\documentclass[sigconf]{acmart}

\usepackage{color}
\usepackage{algorithm}
\usepackage{algorithmic}
\usepackage{multirow}
\usepackage{multicol}
\usepackage{listings}
\usepackage{xcolor}
\usepackage{subfigure}
\usepackage{xspace}

\newcommand{\approach}{CroCS\xspace}

\AtBeginDocument{%
  \providecommand\BibTeX{{%
    \normalfont B\kern-0.5em{\scshape i\kern-0.25em b}\kern-0.8em\TeX}}}

\copyrightyear{2022}
\acmYear{2022}
\setcopyright{acmcopyright}\acmConference[ICSE '22]{44th International Conference
on Software Engineering}{May 21--29, 2022}{Pittsburgh, PA, USA}
\acmBooktitle{44th International Conference on Software Engineering (ICSE '22), May
21--29, 2022, Pittsburgh, PA, USA}
\acmPrice{15.00}
\acmDOI{10.1145/3510003.3510125}
\acmISBN{978-1-4503-9221-1/22/05}

\begin{document}

\title{Cross-Domain Deep Code Search with Meta Learning}
\author{Yitian Chai$^1$, Hongyu Zhang$^2$, Beijun Shen$^1$, Xiaodong Gu$^1$}
\authornote{Xiaodong Gu is the corresponding author}
\affiliation{%
  \institution{$^1$School of Software, Shanghai Jiao Tong University, China}
}
\affiliation{%
  \institution{$^2$The University of Newcastle, Australia}
}
\email{{sjtu_chaiyt,bjshen, xiaodong.gu}@sjtu.edu.cn, hongyu.zhang@newcastle.edu.au}

\renewcommand{\shortauthors}{Chai, et al.}


\definecolor{dkgreen}{rgb}{0,0.6,0}
\definecolor{gray}{rgb}{0.5,0.5,0.5}
\definecolor{mauve}{rgb}{0.58,0,0.82}
\lstset{frame=tb,
     language=SQL,
     aboveskip=3mm,
     belowskip=3mm,
     showstringspaces=false,
     columns=flexible,
     basicstyle = \ttfamily\small,
     numbers=none,
     numberstyle=\tiny\color{gray},
     keywordstyle=\color{blue},
     commentstyle=\color{dkgreen},
     stringstyle=\color{mauve},
     breaklines=true,
     breakatwhitespace=true,
     tabsize=1
}

\begin{abstract}

 Recently, pre-trained programming language models such as CodeBERT have demonstrated substantial gains in code search. Despite showing great performance, they rely on the availability of large amounts of parallel data to fine-tune the semantic mappings between queries and code. This restricts their practicality in domain-specific languages that have relatively scarce and expensive data. 
 In this paper, we propose \approach, a novel approach for domain-specific code search. \approach employs a transfer learning framework where an initial program representation model is pre-trained on a large corpus of common programming languages (such as Java and Python), and is further adapted to domain-specific languages such as \emph{Solidity} and \emph{SQL}. Unlike cross-language CodeBERT, which is directly fine-tuned in the target language, \approach adapts a few-shot meta-learning algorithm called MAML to learn the good initialization of model parameters, which can be best reused in a domain-specific language. 
 We evaluate the proposed approach on two domain-specific languages, namely Solidity and SQL, with model transferred from two widely used languages (Python and Java).
 Experimental results show that \approach significantly outperforms conventional pre-trained code models that are directly fine-tuned in domain-specific languages, and it is particularly effective for scarce data. 
\end{abstract}



\maketitle

	\section{Introduction}
	
	Recently, deep neural networks (DNN) have been widely utilized for code search~\cite{gu2018deepcs, sachdev2018retrieval, feng2020codebert, husain2018create, cambronero2019deep, YanYCSJ20}. Unlike traditional keyword matching methods~\cite{489076, lange2008swim, bajracharya2006sourcerer, 2015CodeHow, lu2015query, lemos2014thesaurus}, deep code search models employ deep neural networks to learn the representations of both queries and code, and measure their similarities through vector distances. The application of DNNs significantly improves the understanding of code semantics, thereby achieving superb performance in code search tasks~\cite{gu2018deepcs,feng2020codebert,yao2019coacor, LiQYSC20}.
	
	A major challenge for deep code search is the adaptation of deep learning models to domain-specific languages. 
	State-of-the-art code search methods are mainly designed for common languages such as Java and Python. They rely heavily on the availability of large parallel data to learn the semantic mappings between code and natural language~\cite{fu2017easy}. 
	On the other hand, there is an emerging trend of domain-specific languages such as Solidity for smart contracts~\cite{zakrzewski2018towards, wohrer2018smart, yang2021multi} 
	where code search is also needed. There is often insufficient training data in specific domains, causing poor fit of deep learning models.
	Furthermore, for each specific domain, the costs of data collection, cleaning, and model training for constructing an accurate model are all non-neglectable. 


    One potential route towards addressing this issue is the pre-trained code models, which pre-train a common representation model on a large, multilingual code corpus, and then fine-tune the model on task-specific data~\cite{salza2021effectiveness}. This enables code search models to transfer prior knowledge from the data-rich languages to the low-resource language. For example, CodeBERT~\cite{feng2020codebert}, the state-of-the-art code representation model, can be pre-trained on multiple common languages and then fine-tuned in the code search task for a target language~\cite{salza2021effectiveness}. 
	However, it is challenging to reuse knowledge from a mix of source languages for code search in the target language.
	Different languages have their unique characteristics, and correspond to different representations. Parameters learnt from each language can distract each other, resulting in a conflict in the shared representations. 
	This is even more challenging in the domain-specific code search, where the target language usually has scarce training samples. 
	
	In this paper, we present \approach (\textbf{Cro}ss-Domain Deep \textbf{C}ode \textbf{S}earch), a cross-domain code search technique based on few-shot meta learning. \approach extends the ``pretraining-finetuning'' paradigm of CodeBERT with a meta learning phase that explicitly adapts the model parameters learnt from multiple source languages to the target language.  
	\approach begins by pre-training CodeBERT on a large corpus of multiple common languages such as Java and Python.
	Then, a meta learning algorithm named MAML (Model-Agnostic Meta-Learning) is employed in order to prevent the model parameters from falling into the local optimization of source languages. 
	The goal of this algorithm is to find the initialization of model parameters that enables fast adaptation to a new task with a small amount of training examples. 

	To evaluate the effectiveness of \approach, we pre-train \approach on a large corpus of common languages such as Python and Java. Then, we perform code search on two domain-specific datasets written in Solidity and SQL. 
	We compare our approach with three baseline models, namely, a neural code search model without pre-training, a within-domain pre-training model CodeBERT~\cite{feng2020codebert}, and a cross-language CodeBERT~\cite{salza2021effectiveness} that directly fine-tunes the target language on a pre-trained model. 
    Experimental results show that \approach significant outperforms within-domain counterparts. In particular, our approach shows more strength when the data is scarce, indicating the superb effectiveness of our approach in cross-domain code search.   	

	The contributions of this work can be summarized as:
	 \begin{itemize}
	     \item We propose \approach, a novel cross-domain code search method using few-shot meta learning.
	     \item We extensively evaluate \approach on a variety of cross-language code search tasks. Experimental results have shown that \approach outperforms the pre-training and fine-tuning counterparts by a large margin. 
	 \end{itemize}

	\section{Background}
	
	\subsection{Code Search Based on Deep Learning}
	The past few years have witnessed a rapid development of deep learning for software engineering, in which code search has been one of the most successful applications. 
	Compared with traditional text retrieval methods, deep learning based code search learns representations of code and natural language using deep neural networks, and thus has achieved superb performance~\cite{gu2018deepcs,feng2020codebert,sachdev2018retrieval, husain2018create, cambronero2019deep}.
	
	\begin{figure}[htbp]
        \includegraphics[scale = 0.3]{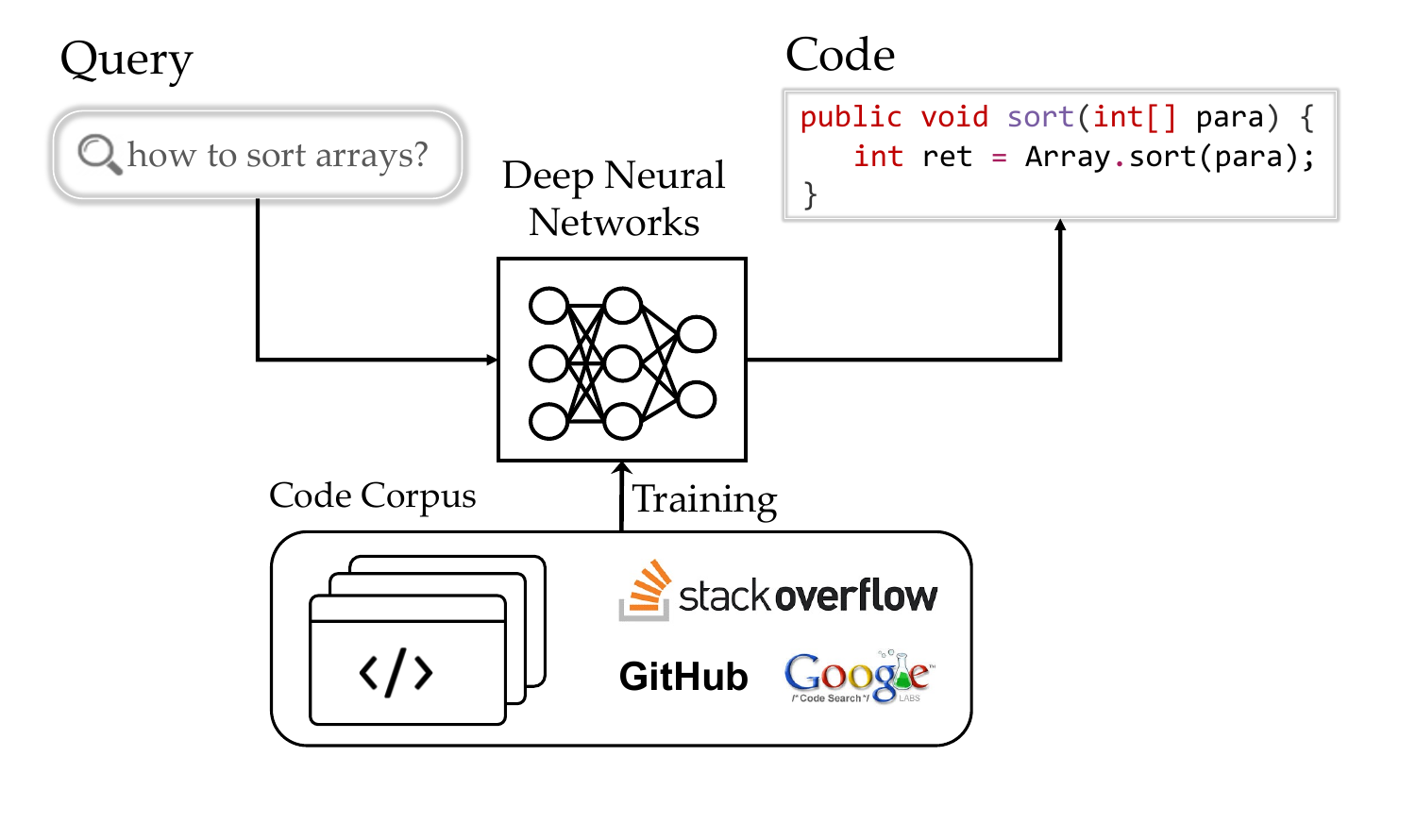}
        \caption{Deep learning based code search.}
        \label{fig1}
    \end{figure}
    
	Figure~\ref{fig1} shows the overall framework of deep learning based code search. In the training phase, a bi-modal deep neural network is trained based on a large parallel corpus of code and natural language to learn the semantic representations (high-dimensional vectors) of both queries and code snippets. Then, a similarity function is employed to numerically compute the similarity between code and query vectors. The model is usually trained by minimizing the triplet ranking loss~\cite{reimers2019sentencebert}, namely,
	\begin{equation}
	    \mathcal{L}(c, \mathbf{d}+, \mathbf{d}-) = \mathrm{max}(\mathrm{cos}(\mathbf{c},\mathbf{d}+)-\mathrm{cos}(\mathbf{c},\mathbf{d}-)+\epsilon, 0)
	\end{equation}
	where $\mathbf{c}$, $\mathbf{d}+$, and $\mathbf{d}-$ represent the vector representations for the code, the correct description, and the distracting description, respectively. \emph{cos} denotes the cosine similarity between two vectors. $\epsilon$ is a margin which ensures that $\mathbf{d}+$ is at least $\epsilon$ closer to $\mathbf{c}$ than $\mathbf{d}-$~\cite{reimers2019sentencebert}.
	
	In the search phase, the search engine is given a query from the user. It computes the vectors for both the query and code snippets in the codebase using the trained model. Then, it goes through the codebase and matches the query with each code snippet according to their vector distances. Snippets that have the best matching scores are returned as the search results.

	\subsection{Pre-trained Models for Code Search}
	
	
	Recently, pre-trained models such as BERT~\cite{devlin2019bert} and GPT-2~\cite{radford2019language} have achieved remarkable success in the field of NLP~\cite{devlin2019bert,radford2019language}. 
	As such, researchers start to investigate the adaptation of pre-trained models to software programs~\cite{feng2020codebert,,wang2021codet5}. Code search is one of the most successful applications of pre-trained models for programming languages.
	
\begin{figure}
	    \centering
        \centering
        \includegraphics[scale = 0.3]{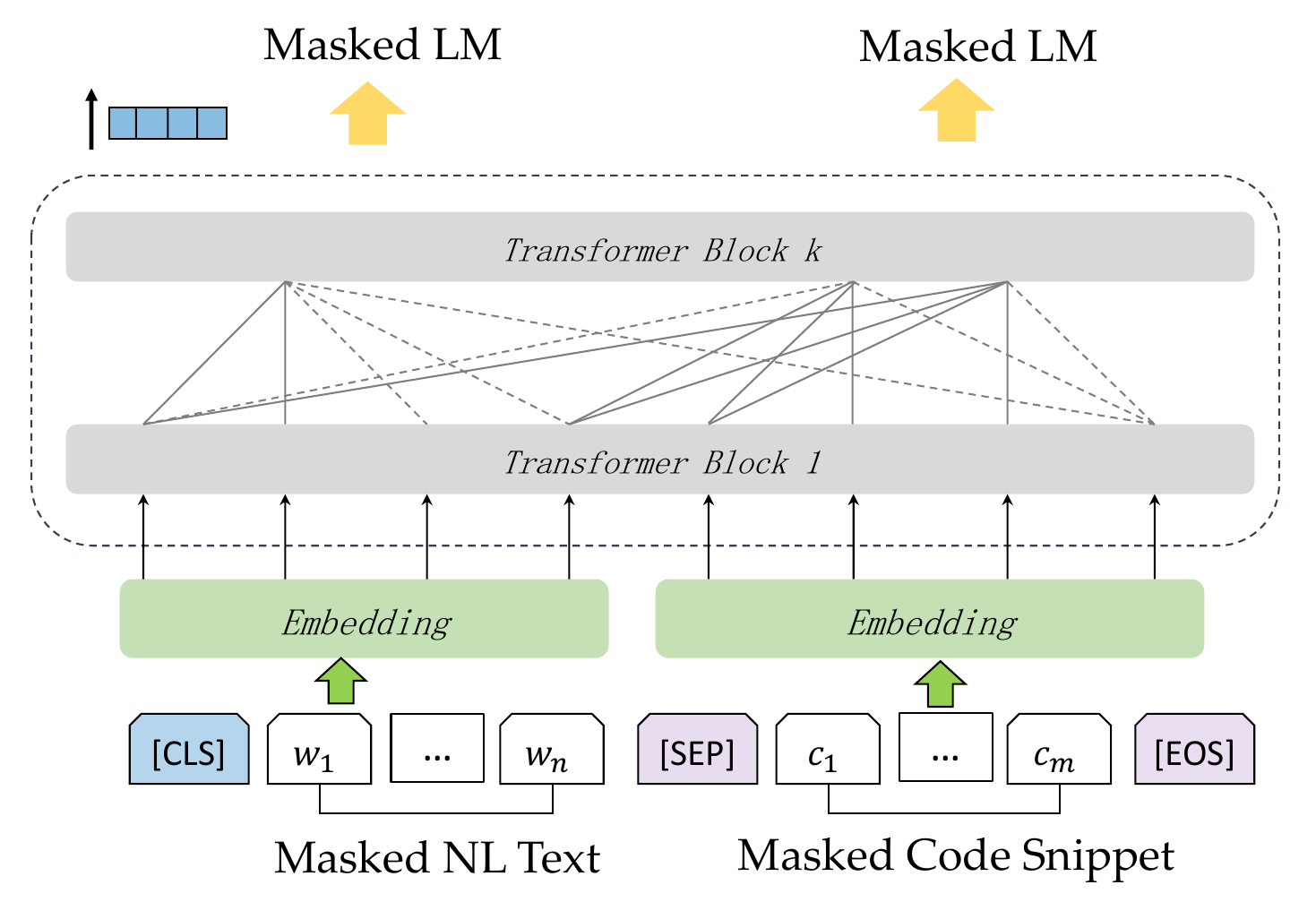}
	    \caption{Architecture of CodeBERT with masked language model.}
	    \label{fig:archBERT}
	\end{figure}

	One of the most successful pre-trained models for code is the CodeBERT~\cite{feng2020codebert}. CodeBERT is built on top of BERT~\cite{devlin2019bert} and Roberta~\cite{liu2019roberta}, two popular pre-trained models for natural language. Unlike pre-trained models in NLP, CodeBERT is designed to represent bi-modal data~\cite{casalnuovo2018studying}, namely, programming and natural languages.
	Figure~\ref{fig:archBERT} shows the architecture of CodeBERT. In general, the model is built upon a \emph{Transformer} encoder. The training involves two pre-training tasks in six programming languages. One is the masked language modeling (MLM), which trains the model to fill the masked token in the input sequences. The other task is the replaced token detection (RTD), which trains the model to detect the replaced tokens in the input sequences. 
	These two pre-training tasks endow CodeBERT with generalization ability, so that it can be fine-tuned to adapt to downstream tasks such as code search and code summarization.

	As a code representation model, CodeBERT has been successfully employed for code search~\cite{feng2020codebert}. Specifically, a binary classifier is employed which takes as input the representation of the [CLS] token and predicts whether a given <NL, PL> pair is semantically related. This classifier is then fine-tuned on a code search dataset by minimizing the cross-entropy loss. In the search phase, the classifier predicts the matching score between an NL query and each code snippet in the codebase. The search engine returns the top-$k$ code snippets that have the highest matching scores. 

	Due to the superb performance, researchers have also applied CodeBERT for cross-language code search~\cite{salza2021effectiveness}. 
	They pre-trained CodeBERT with multiple languages such as Python, Java, PHP, Javascript, and Go, and then fine-tuned a code search model on an unseen language such as Ruby. 
	Results have shown that cross-language code search achieves better performance than training in a single language from scratch. This further supports the effectiveness of transfer learning for code search~\cite{salza2021effectiveness}.

	\subsection{Meta Learning and Few-Shot Learning} 

	Few-shot learning is a machine learning technology that aims to quickly adapt a trained model to new tasks with less examples~\cite{snell2017prototypical}. 
	Despite the superb performance, deep learning models are often data-hungry~\cite{fu2017easy}. They rely on the availability of large-scale data for training. That means, the performance can be limited due to the scarcity of data in specific domains~\cite{fu2017easy}. By contrast, humans can learn knowledge from a few examples. For example, a child can learn to distinguish between lions and tigers when provided with a few photos, probably because human beings have prior knowledge before learning new data or because human brains have a special way to process knowledge. 
	Based on this intuition, researchers have proposed few-shot learning. 
	
	Few-shot learning methods can be roughly classified into the following two categories:\\
	1) \textbf{Metric-based methods}, which learn a distance function between data points so that new test samples can be classified through comparison with the $K$ labeled examples~\cite{yin2020metalearning}. 
	There are a few typical algorithms for metric-based few-shot learning, such as \emph{Siamese Network}~\cite{chopra2005learning}, \emph{Prototypical Network}~\cite{snell2017prototypical}, and \emph{Relation Network}~\cite{sung2018learning}.
	    
	    

	\noindent 2) \textbf{Meta Learning}, also known as ``learning-to-learn'', which trains a model on a variety of learning tasks, such that it can solve new learning tasks using only a small number of training samples~\cite{finn2017maml}. Unlike the conventional machine learning prototype that a model is optimized in the training set to minimize the training loss, meta learning updates model parameters using the validation loss in order to enhance the generalization to different tasks. 
	There are some typical algorithms for few-shot meta learning, such as \emph{MAML}~\cite{finn2017maml} and \emph{Reptile}~\cite{nichol2018first}.

	MAML (Model-Agnostic Meta-Learning) is a few-shot meta learning algorithm which aims at learning a good initialization of model parameters so that the model can quickly reach the optimal point in a new task with a small number of data samples~\cite{yin2020metalearning,finn2017maml}.  
	The algorithm assumes that the data used for training follows a distribution $p(T)$ over $k$ tasks $\{T_1,...,T_k\}$, where $T_i$ stands for a specific machine learning task on the data.
    The intuition is that some data features are more transferrable than others. In other words, they are broadly applicable to all tasks in $p(T)$, rather than a single individual task $T_i$. To find such general-purpose representations, MAML updates model parameters that are sensitive to changes in the task, such that small changes in the parameters will produce large improvements on the loss function of any task drawn from $p(T)$. Motivated by this, MAML separates data into individual tasks. A meta learner is employed to update parameters using gradients on each local task~$T_i$~\cite{finn2017maml}. A more detailed description of the algorithm and how it is adapted to code search will be presented in Section~\ref{ss:approch:meta}. 

	\section{Approach}
	
    
	\subsection{Overview}\label{AA}
\begin{figure}[tb]
        \includegraphics[scale = 0.24]{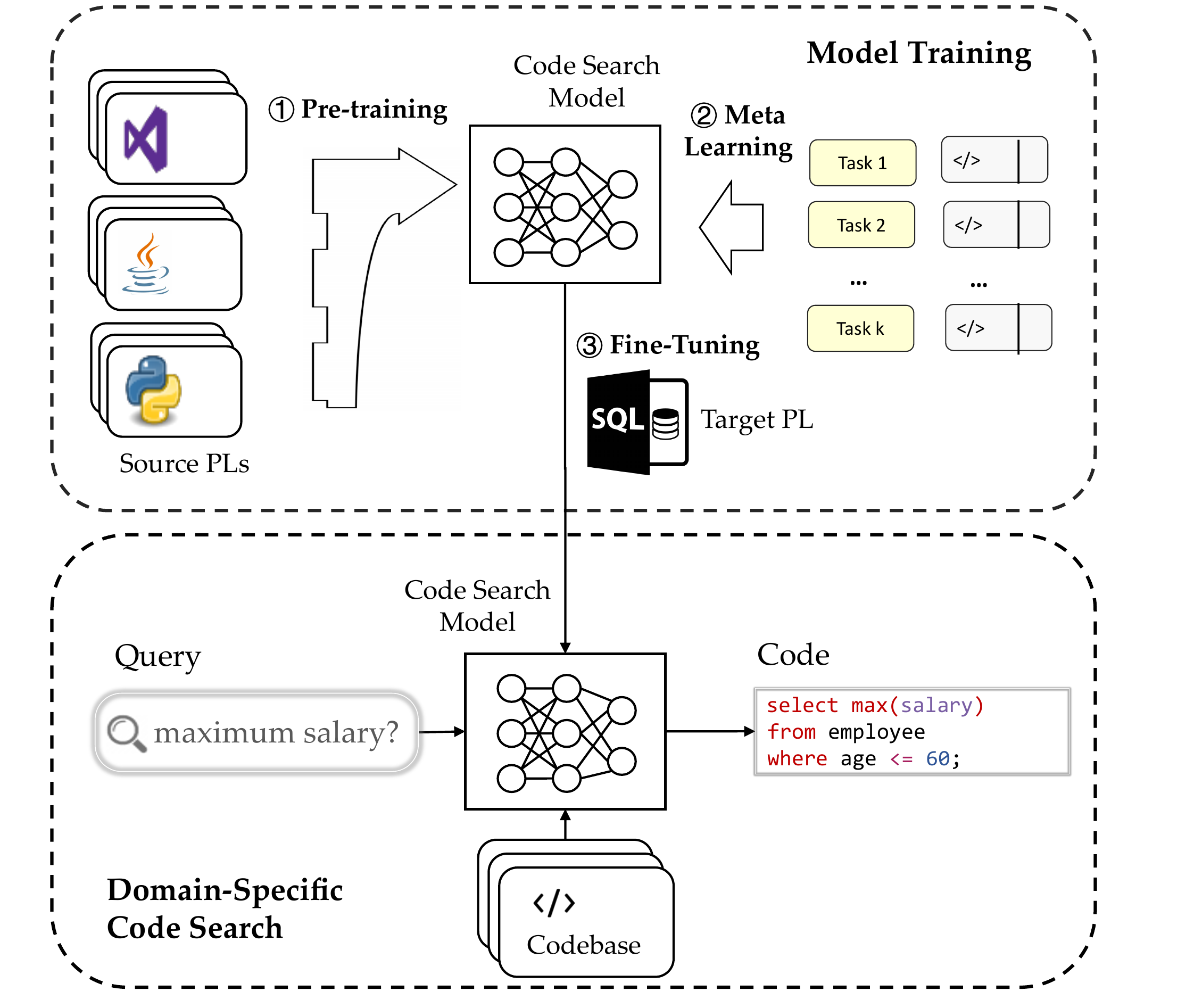}
        \caption{Architecture of \approach.}
        \label{fig3}
    \end{figure}

	Figure~\ref{fig3} shows the architecture of \approach. In general, \approach takes CodeBERT~\cite{feng2020codebert} as the backbone, and extends it with a meta learning phase. 
	The core component of \approach is RoBERTa~\cite{liu2019roberta}, which is built upon a multi-layer bidirectional Transformer~\cite{vaswani2017attention} encoder. 
	
    The pipeline of \approach involves four phases. 
	Similar to CodeBERT, we start by pre-training \approach to learn code representations in a large corpus of multiple source languages. 
	Next, we perform meta learning to explicitly transfer the representations of source languages into the target language. 
	After the domain adaptation, we fine-tune it on the code search data of the target language in order to train the semantic mapping between code and natural language. We finally perform code search using the fine-tuned model. 
	We will describe the detailed design of each phase in the following sections.

	\subsection{Pre-training}
	
	The pre-training phase aims to learn code and NL representations from a large corpus of multiple common languages such as Java and Python. Similar to CodeBERT, we use the pre-training task of masked language modeling (MLM). We did not use the RTD (replaced token detection) pre-training task of CodeBERT because the effect of this task has been shown to be marginal~\cite{feng2020codebert}.
	
   In the pre-training phase, the model takes as input an $\langle$NL, PL$\rangle$ pair which is formatted into a sequence of $$[CLS], w_1, w_2,... w_n, [SEP], c_1, c_2,..., c_m, [EOS]$$ where $w_1, w_2, ..., w_n$ denotes a sequence of $n$ words in the natural language text, while $c_1, c_2, ..., c_m $ represents a sequence of $m$ tokens in the code snippet. 
   The special $[CLS]$ token at the beginning is a placeholder for the representation of the entire input sequence. The $[SEP]$ token indicates the border of the code snippet and the natural language text. The $[EOS]$ token indicates the end of the sequence. 
   
   During the pre-training process, we randomly replace 15\% of the tokens in the input sequence with a special $[MASK]$ token and let the model predict the original token. The task can be optimized by minimizing the cross-entropy loss between the predicted and the original tokens. 
   	
   The pre-trained model can be used to produce the contextual vector representation of each token for both natural language descriptions and code snippets. In particular, the representation of the $[CLS]$ token stands for the aggregated sequence representation which can be used for classifying the entire input sequence.

	\subsection{Meta Learning}\label{ss:approch:meta}

	We next perform meta learning to adapt the pre-trained code model to the target domain. We employ a meta-learning algorithm named MAML (Model-Agnostic Meta-Learning)~\cite{finn2017maml} which is a typical algorithm for few-shot learning~\cite{finn2017maml,gu2018meta,sun2019meta}. 
	The key idea of MAML is to use a set of source tasks \{$T_1$,$\ldots$, $T_k$\} to find the initialization of parameters $\theta_0$ from which learning a target task $T_0$ would require only a small number of training samples~\cite{finn2017maml}. In the context of code search, this amounts to using large data of common languages to find good initial parameters and training a new code search model on a small, domain-specific language starting from the found initial parameters.
	We formulate code search as a binary classification task $T$ which predicts whether a given $\langle NL, PL\rangle$ pair matches (1 = match, 0 = irrelevant). Unlike CodeBERT which directly fine-tunes on the code search task~$T$, the MAML algorithm assumes that the dataset used for training follows a distribution $p(T)$ over $k$ tasks $\{T_1,...,T_k\}$. Hence, it splits $T$ into a set of $k$ tasks~$\{T_1,...,T_k\}$.
	Each task~$T_i$ aims at training a code search model with small sized data, therefore simulates the low-resource learning. 
	Based on this idea, each $T_i$ is assigned to train the code search model in a private training and validation set denoted as $T_i$ $\sim$ $\{D_{\mathrm{train}}, D_{\mathrm{valid}}\}$. 
	
	Let $\theta$ denote the global parameters for the entire model and $\theta_i$ denote the local parameters for task~$T_i$. A meta learner is trained to update model parameters $\theta_i$ using one or more gradient descent updates on task $T_i$. For example, when using one gradient update, the training step can be formulated as
	\begin{equation}
		\label{eq1}
		\theta_{i}=\theta-\alpha\nabla_{\theta}L_{T_i}(f_\theta),\quad i=1,\ldots,k
	\end{equation}
	where $f_\theta$ denotes the deep learning model for specific task with parameters~$\theta$; $L_{T_i}$ represents the loss function for task~$T_i$; $\alpha$ denotes the step size for each task and is fixed as a hyperparameter for the meta learner. 
	
	In our approach, the training set~$D^{\mathrm{train}}$ (involves multiple source languages) for the original code search task~$T$ is randomly segmented into $k$ batches~$\{D_1,...,D_k\}$ equally. Each $D_i$ is used as the data set for the local task~$T_i$. 
	To perform the local task, $D_i$ is further split into a training and validation set $\{D_i^\mathrm{train}, D_i^\mathrm{valid}\}$ with the same data size. 
	Each $T_i$ is then performed on $\{D_i^\mathrm{train}, D_i^\mathrm{valid}\}$ to obtain the local gradient $\nabla_{\theta}L_{T_i}(f_\theta)$. These local gradients are aggregated by the meta-learner every $M$ steps in order to update the global parameter~$\theta$.
	
	In order to learn a good model initialization of multiple source languages, we construct the D$^\mathrm{train}$ from multiple source languages. We segment the original dataset of each language into batches. This results in a pool of batches that involves multiple languages. During meta learning, we randomly select $k$ batches from the batch pool. 

	The procedure of our algorithm is summarized in Algorithm~1.
	
	\begin{figure}[htbp]
        \includegraphics[scale = 0.4]{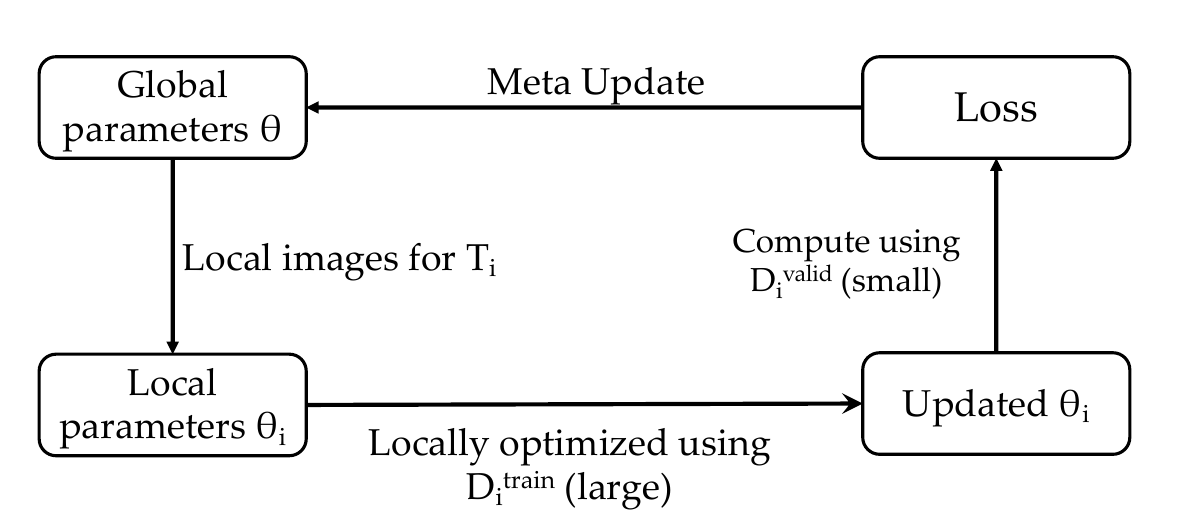}
        \caption{An overview of the MAML algorithm.}
        \label{fig:maml}
    \end{figure}
    
    \begin{algorithm}
        \caption{Meta Learning for Code Search}
        \begin{algorithmic}[1]
        \REQUIRE $\alpha$, $\beta$: step size; $M$: meta update steps
        \vspace{1mm}
        \STATE Pre-train the global model on source languages and obtain the initial parameters~$\theta$
        \STATE Create $k$ copies of $\theta$ with each $\theta_i$ being the local parameters for $T_i$.
            \WHILE{not done}
                \STATE Divide the dataset of each source language into batches
                \STATE Construct $D^\mathrm{train}$ by randomly selecting $k$ batches from the batch pool, with the $i$-th batch~$D_i$ assigned for task $T_i$
                \FOR {\textbf{each} $D_i\in D^\mathrm{train}$}
                    \STATE Split $D_i$ into $\{D_i^\mathrm{train}, D_i^\mathrm{valid}\}$
                    \STATE Run $T_i$ on $\{D_i^\mathrm{train}, D_i^\mathrm{valid}\}$ and evaluate local gradients $\nabla_{\theta}L_{T_i}(f_\theta)$ using the cross-entropy loss~$L_{T_i}$
                    \STATE Update local parameters~$\theta_i$ with gradient descent:\\ \quad\quad$\theta_{i}=\theta-\alpha\nabla_{\theta}L_{T_i}(f_\theta)$
                    \IF {$i$ mod $M$ == 0}
                        \STATE Evaluate gradients $\nabla_{\theta}L_{T_i}(f_{\theta_{i}})$ using the cross-entropy loss $L_{T_i}$ in $D_i^\mathrm{valid}$
                        \STATE Update the global parameters~$\theta$ using the gradients on the validation set:\\ \quad\quad $\theta \Leftarrow \theta-\beta\nabla_{\theta}L_{T_i}(f_{\theta_{i}})$ 
                    \ENDIF
                \ENDFOR
            \ENDWHILE
        \end{algorithmic}
    \end{algorithm}
	
	\subsection{Fine-Tuning}
	
	In the fine-tuning phase, we adapt \approach to the code search task in the target language. 
    We fine-tune the model on the code search task, which can be formulated as a binary classification problem. For a corpus of $\langle NL, PL\rangle$ pairs, we create the same number of negative samples by randomly replacing NL or PL in the original pairs. We assign a label to each pair to indicate whether the NL is corresponding to the PL in the pair (1=relevant, 0 =irrelevant). 
    
    For each training instance, we build an input sequence with the same format as in the pre-training phase. We take the hidden state in the $[CLS]$ position of CodeBERT as the aggregated representation of the input sequence. The representation is further taken as input to a fully connected neural classifier to predict whether the given $\langle NL, PL\rangle$ pair is relevant. We fine-tune the model by minimizing the binary cross-entropy loss between predictions and labels.

    \subsection{Domain-Specific Code Search}
    Finally, we perform code search based on the fine-tuned model in a domain-specific codebase. The code search engine works with the following steps:
    \begin{itemize}
        \item[1)] A natural language query~$Q$ is provided to the code search system.
        \item[2)] Splice Q separately with each code snippet $C_i$ in the codebase to obtain a series of input sequences $$<Q, C_1>, \ldots, <Q, C_n>$$
        \item[3)] Input these sequences into the trained model and obtain their matching scores.
        \item[4)] Sort code snippets according to their matching scores. 
        \item[5)] Return the top-k code snippets as the results.
    \end{itemize}

	\section{Experimental Setup}
	We evaluate the performance of \approach in domain-specific code search tasks and explore the effect of training data size on the performance. Finally, we extend our method to other backbone pre-trained models such as GPT-2~\cite{radford2019language}. 
	In summary, we evaluate \approach by addressing the following research questions:
	
	\begin{itemize}
	    \item \textbf{RQ1: How effective is \approach in cross-domain code search?}
	    
	To verify the effectiveness of \approach in cross-domain code search tasks, we take Python and Java as the source languages and adapt the learned model to two domain-specific languages, namely, Solidity and SQL. We compare the accuracy of code search by various approaches in the two target languages. 

	\item \textbf{RQ2: What is the impact of data size on the performance of cross-domain code search? }
	
	As mentioned, one of the challenges for cross-domain code search is the scarcity of data in the domain-specific language. In RQ2, we aim to study the effect of data size on the performance. We vary the size of dataset 
	and compare the performance under different data sizes. 



	

    \item \textbf{RQ3: How effective is \approach applied to other pre-trained programming language models?} 
	
    Besides CodeBERT, there are other pre-trained models that also achieve outstanding results in software engineering tasks~\cite{mastropaolo2021t5code,ahmad2021unified,phan2021cotext}. We wonder whether other pre-trained models can have the same effectiveness on code search when equipped with meta learning. We replace the backbone pre-trained model with GPT-2~\cite{radford2019language, brown2020language}, which is also a popular pre-trained language model based on Transformer. GPT-2 differs from BERT in that it is an autoregressive language model built on top of the Transformer decoder. 
	We evaluate the effectiveness of $\mathrm{\approach}_\mathrm{GPT-2}$ and compare it with those of baseline models.
	
	\item \textbf{RQ4: How do different hyperparameters affect the performance of \approach?}
	
	In order to study the effect of hyperparameters to the performance of \approach, we assign different hyperparameters to \approach 
	and examine their impact to the performance of code search.  

    \end{itemize}

	\subsection{Implementation Details}
	We build our models on top of the RoBERTa~\cite{liu2019roberta} using the same configuration as RoBERTa-base (H=768, A=12, L=12). The rate of masked tokens is set to 15\%. 
	We use the default CodeBERT tokenizer, namely, \emph{Microsoft/codebert-base-MLM} with the same vocabulary size (50265). We set the maximum sequence length to 256 to fit our maximum computational resources. The default batch size is set to 64. The three hyperparameters $\alpha, \beta, M$ in Algorithm 1 are empirically set to 1$e$-5, 1$e$-4, and 100, respectively.
	Our experimental implementation is based on the tool provided by Huggingface Transformers\footnote{https://huggingface.co/transformers/} and the higher library provided by Facebook Research\footnote{https://higher.readthedocs.io/}. 
	
	All models are trained on a GPU machine with Nvidia Tesla V100 32G using the Adam~\cite{kingma2017adam} algorithm.
	We use a learning rate of 5$e$-5~\cite{feng2020codebert} in the pre-training phase which warms up in the first 1,000 steps and linearly decays. 
	We measure the performance on the validation set during the training process, and select the checkpoint of the model which has the best accuracy on the validation set for testing.

	\subsection{Datasets}
	    \subsubsection{Data Used for Pre-training and Meta Learning}
		We pre-train and perform meta learning using the training data for the code search task provided by CodeBERT~\cite{feng2020codebert}. We select two popular languages, namely, Python and Java as the source languages. 
		The statistics of the dataset are shown in Table~\ref{table_data}. For each language, the dataset contains parallel data of $\langle$NL, PL$\rangle$ pairs, including both positive and negative samples. In order to prevent the training from falling into a local optimum of one source language, we use only positive samples for pre-training and use the entire set of pairs for meta learning.
	
	
	\begin{table}[!t]  
		\centering
		\caption{Statistics of datasets for pre-training and meta learning.}
		\label{table_data}
		\begin{tabular}{llll}  
			\toprule   
			 Phase &  & Python & Java\\  
			\midrule   
		    pre-train & \# functions & 412,178 & 454,451  \\
		             & \# comments & 412,178 & 454,451 \\
		    \hline
			meta learning & \# functions & 824,342 & 908,886  \\
		    	     & \# comments  & 824,342 & 908,886 \\
			\bottomrule  
		\end{tabular}
	\end{table}
	
	\begin{table}[!t]  
		\centering
		\caption{Number of functions on the dataset of target languages.}
		\label{table_ft}
		\begin{tabular}{lcll}  
			\toprule   
			Language &Train (Finetune) & Valid & Test   \\  
			\midrule
			Solidity & 56,976 & 4,096 & 1,000 \\ 
			\midrule   
			SQL & 14,000 & 2,068 & 1,000 \\		
			\bottomrule  
		\end{tabular}
	\end{table}
	
	\subsubsection{Data Used for Fine-tuning and Code Search}
    We fine-tune and test the code search task using two domain-specific languages, namely, Solidity and SQL~\cite{yang2021multi}. The statistics about the datasets are shown in Table~\ref{table_ft}. 
    
    Solidity is an object-oriented language that is specifically designed for smart contracts~\cite{yang2021multi}. 
	The dataset of Solidity used in our experiments is provided by~\cite{yang2021multi} for smart contract code summarization. We preprocess the dataset by removing all inline comments from functions. We remove duplicate pairs, namely, two $\langle$NL, PL$\rangle$ pairs that have the same comment but differ only in the number of position in the dataset and a few variable names in code. We also balance positive and negative samples where the negative samples are generated by randomly replacing NL (i.e. ($\mathbf{c}$, $\hat{\mathbf d}$)) and PL (i.e. ($\hat{\mathbf c}$, $\mathbf{d}$)) of positive samples.
	
	SQL is a well-known language that is specifically designed for manipulating database systems. 
    The dataset we used for fine-tuning and testing SQL is provided by~\cite{Yu&al.18c} for cross-domain semantic parsing and SQL code generation (text-to-SQL). The original data is in a JSON format and contains the following fields:
    \begin{itemize}
        \item question: the natural language question.
        \item question\_toks: the natural language question tokens.
        \item db\_id: the database id to which this question is addressed.
        \item query: the SQL query corresponding to the question.
        \item query\_toks: the SQL query tokens corresponding to the question.
        \item sql: parsed results of this SQL query. 
    \end{itemize}
    
    We preprocess the SQL dataset by selecting the ``question'' and ``query'' fields from the \emph{$.$json} data as our \texttt{NL} and \texttt{PL}, respectively. We remove duplicate data that has the same code from the original test set. We also balance positive and negative samples where the negative samples are generated by randomly disrupting descriptions and code based on positive samples.

	\subsection{Evaluation Metrics}
	We measure the performance of code search using two popular quantitative criteria on the test set, including MRR (Mean Reciprocal Rank) and the top-$k$ accuracy. They are commonly used for evaluating code search engines~\cite{gu2018deepcs,feng2020codebert}. 
	
	\textbf{MRR}~\cite{2015CodeHow, ye2014learning} aims to let a search algorithm score search results in turn according to the search content, and then arrange the results according to the scores in a descend order. 
	For $N$ test queries, the MRR can be computed as
    \begin{equation}
	    MRR=\frac{1}{N}\sum_{i=1}^{N}\frac{1}{Rank(i)}
	\end{equation}
    where Rank(i) represents the position of the correct code snippet in the returned results for query~$i$. 
    The greater the MRR score, the better the performance on the code search task.

    \textbf{Top-k accuracy} measures how many answers in the first $k$ results hit the query. This metric is close to the real-world scenario of search tasks, that is, users want the most matching results to be placed at the top of the results. In our experiments, we compute the top-$k$ accuracy with $k$ = 1, 5, and 10, respectively.
    
     We use the trained model to predict the matching scores of 1,000 $\langle NL, PL\rangle$ pairs in the test set. For each pair, the model computes the similarities between the text description and all 1,000 code snippets. The top-k similar snippets are selected for calculating the evaluation metrics. We report the average score of all the 1,000 pairs in the test set.
    
    \subsection{Comparison Methods}
    We compare our approach with four baseline methods.
    \begin{enumerate}
        \item \textbf{Code Search without Pre-training}, which trains the code search model using only domain-specific data in Table~\ref{table_ft} without pre-training and meta learning. Through comparing to this baseline model, we aim to verify the effectiveness of pre-training and meta learning in our approach.
        
        \item \textbf{Code Search based on pre-trained model with Natural Language}, which fine-tunes the code search model on the domain-specific data in Table~\ref{table_ft} based on the pre-trained model that is initialized by the natural language pre-training models, namely Roberta~\cite{liu2019roberta} and GPT-2~\cite{radford2019language}.

        \item \textbf{Within-domain Code Search with CodeBERT}~\cite{feng2020codebert}, which pre-trains and fine-tunes only with the domain-specific data in Table~\ref{table_ft} without prior knowledge of common languages. 
                
        \item \textbf{Cross-Language Code Search with CodeBERT}~\cite{salza2021effectiveness}, which directly fine-tunes the code search model on the domain-specific data (Table~\ref{table_ft}) on a model that is pre-trained on the data of multiple common languages (Table~\ref{table_data}). Through comparing to this baseline model, we aim to validate the usefulness of meta learning in our approach. 
  
    \end{enumerate}
    
    We implement all baseline models based on the open source code of CodeBERT\footnote{https://github.com/microsoft/CodeBERT} using the same hyperparameters as in the CodeBERT paper~\cite{feng2020codebert}.

	\section{Experimental results}
	
	\subsection{Effectiveness in Cross-Domain Deep Code Search (RQ1)}
	
 	\begin{table}[!t]  
 		\centering
 		\small
 		\caption{Performance of each method in the SQL dataset.}
 		\label{table_sql}
 		\begin{tabular}{lllll}  
 			\toprule   
 			Model &  Acc@1 & Acc@5 & Acc@10 &  MRR   \\  
 			\midrule   
 			No-Pretraining & 0.002 & 0.010 & 0.022 & 0.0124 \\	
 			CodeBERT (NL-based) & 0.652 & 0.926 & 0.966 & 0.7690 \\
 			CodeBERT (within-domain) & 0.607 & 0.899 & 0.945 & 0.7351\\
 			CodeBERT (cross-language) & 0.675 & 0.920 & 0.960 & 0.7818 \\
			
 			\midrule
 			\approach & \textbf{0.746} & \bf 0.952  &\bf 0.972 & \textbf{0.8366} \\ 
 			\bottomrule  
 		\end{tabular}
 	\end{table}
	
 	\begin{table}[!t]  
 		\centering
 		\small
 		\caption{Performance of each method in the Solidity dataset.}
 		\label{table_solidity}
 		\begin{tabular}{lllll}  
 			\toprule   
 			Model &  Acc@1 & Acc@5 & Acc@10 &  MRR   \\  
 			\midrule   
 			No-Pretraing & 0.002 & 0.008 & 0.014 & 0.0101 \\
 			CodeBERT (NL-based) & 0.453 & 0.732 & 0.821 & 0.5801 \\
 			CodeBERT (within-domain) & 0.515 & 0.798 & 0.857 & 0.6383 \\
 			CodeBERT (cross-language) & 0.532 & 0.779 & 0.848 & 0.6436 \\
 			\midrule
 			\approach & \textbf{0.658} & \textbf{0.829} & \textbf{0.879} & \textbf{0.7336} \\ 
 			\bottomrule  
 		\end{tabular}
 	\end{table}

	Table~\ref{table_sql} and \ref{table_solidity} show the performance of different approaches in the cross-domain code search task. We take Python and Java as the source languages and test the performance on two domain-specific languages, namely, SQL and Solidity. 
	
	 \begin{figure*}[htbp]
        \centering
        \subfigure[MRR]{
            \includegraphics[scale = 0.092]{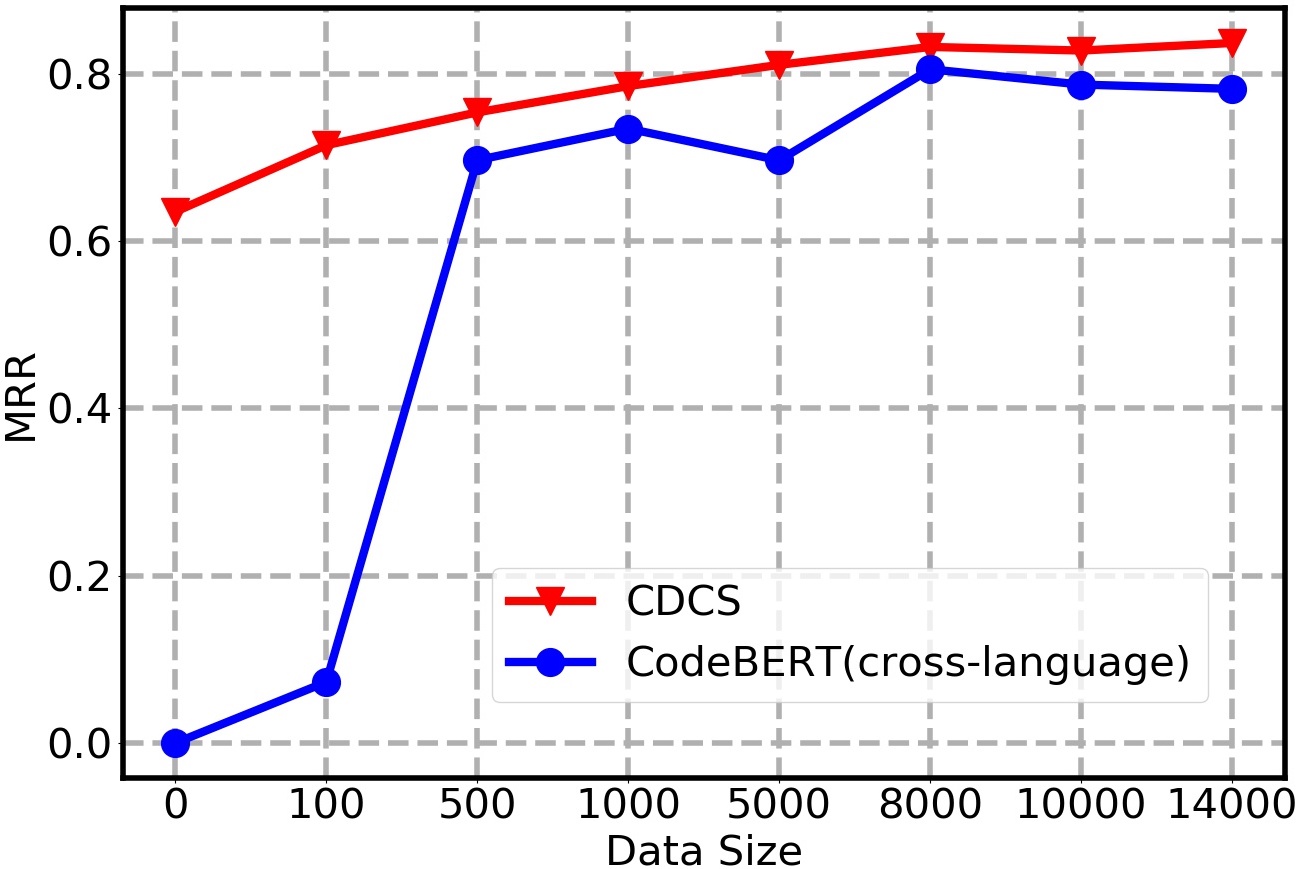}
        }
        \subfigure[Top-1 accuracy]{
            \includegraphics[scale = 0.092]{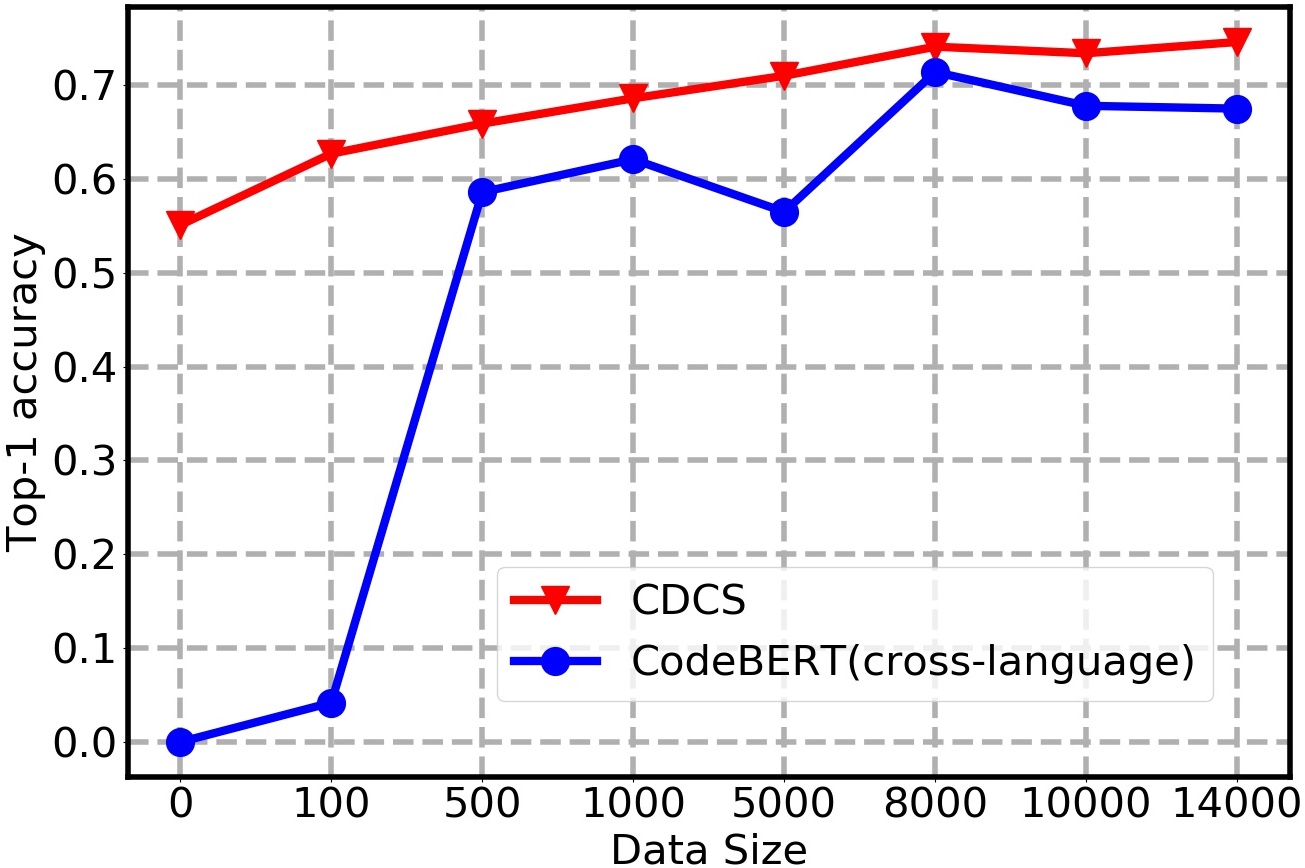}
        }
        \subfigure[Top-5 accuracy]{
            \includegraphics[scale = 0.092]{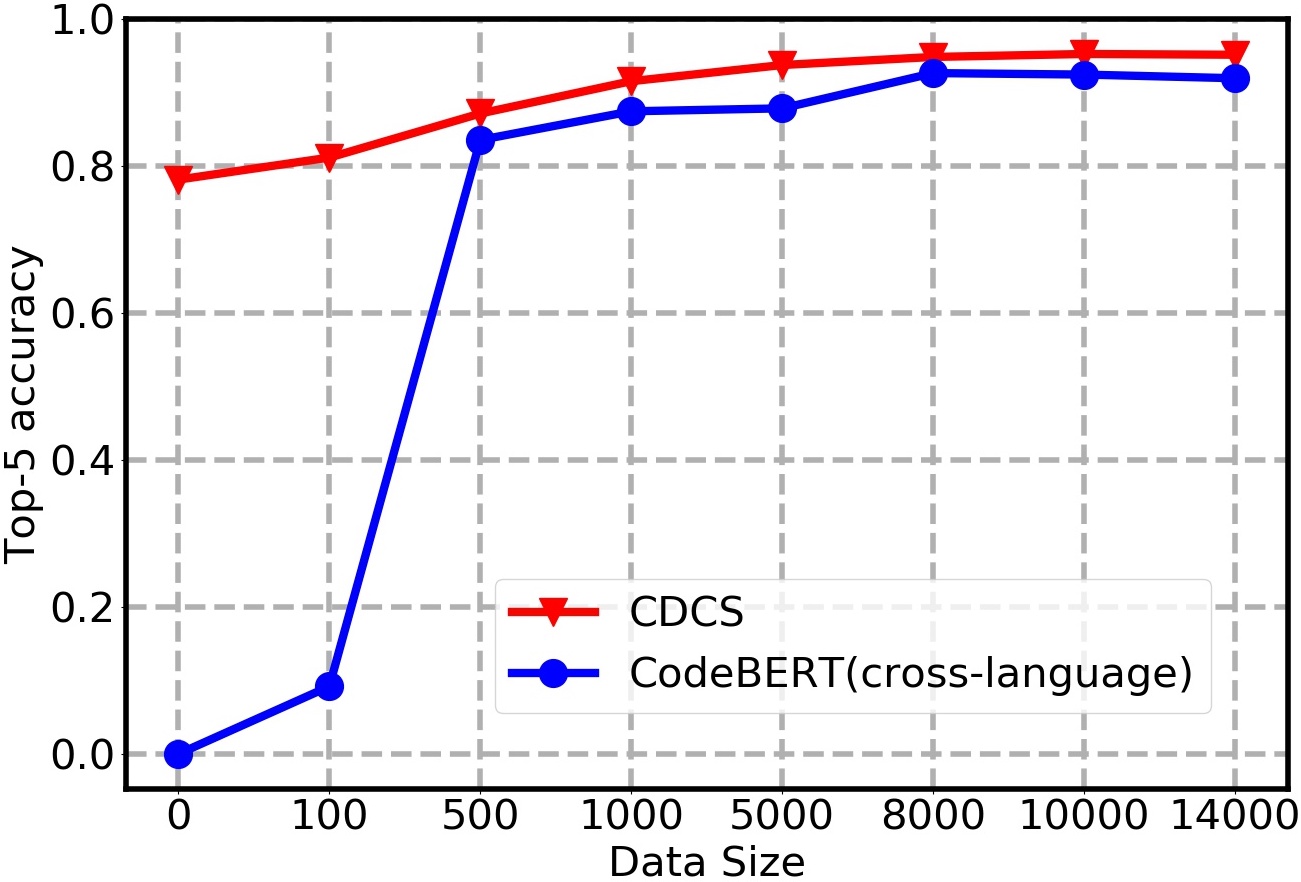}
        }
        \subfigure[Top-10 accuracy]{
            \includegraphics[scale = 0.092]{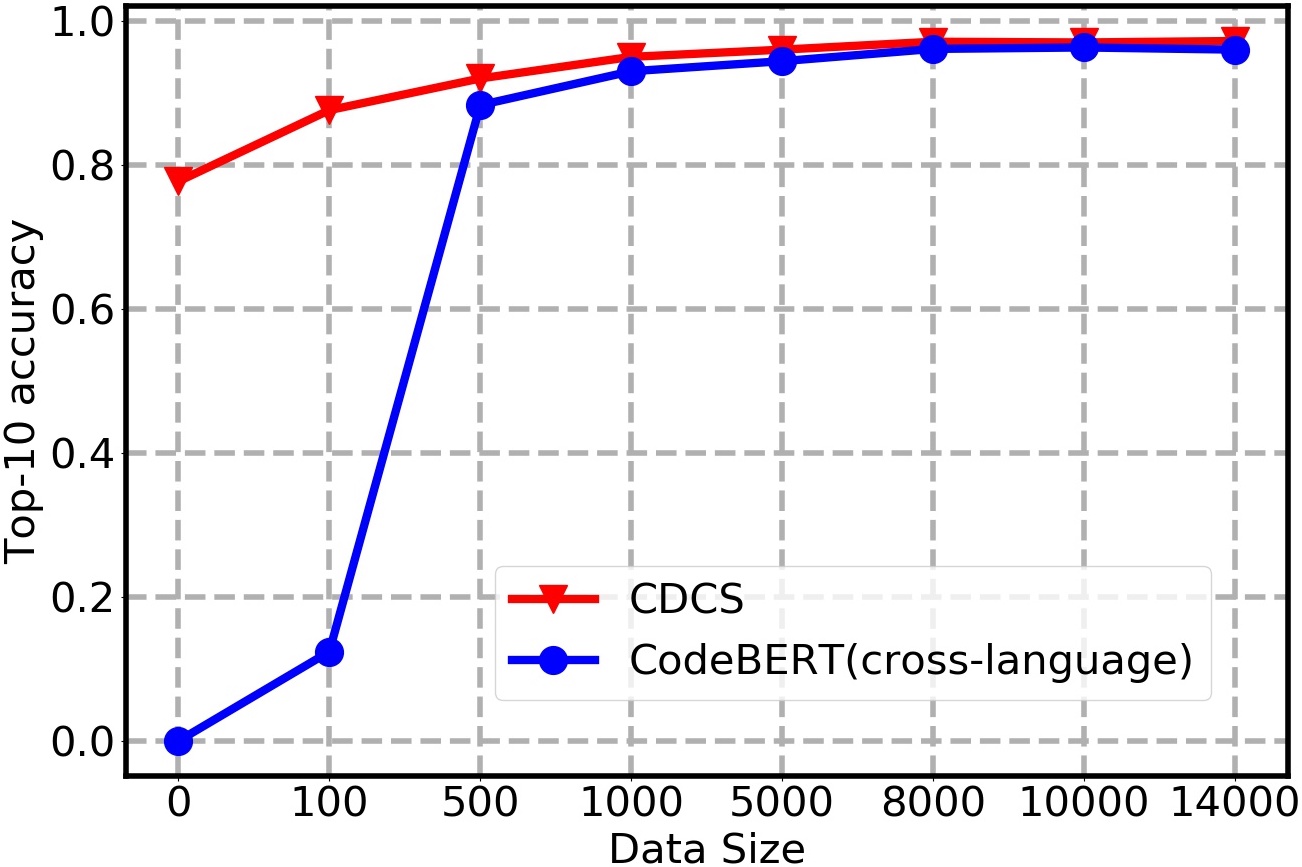}
        }
        \caption{Performance of \approach under different training data sizes on the SQL dataset.}
        \label{RQ2_1}
    \end{figure*}

    \begin{figure*}[htbp]
        \centering
        \subfigure[MRR]{
            \includegraphics[scale = 0.092]{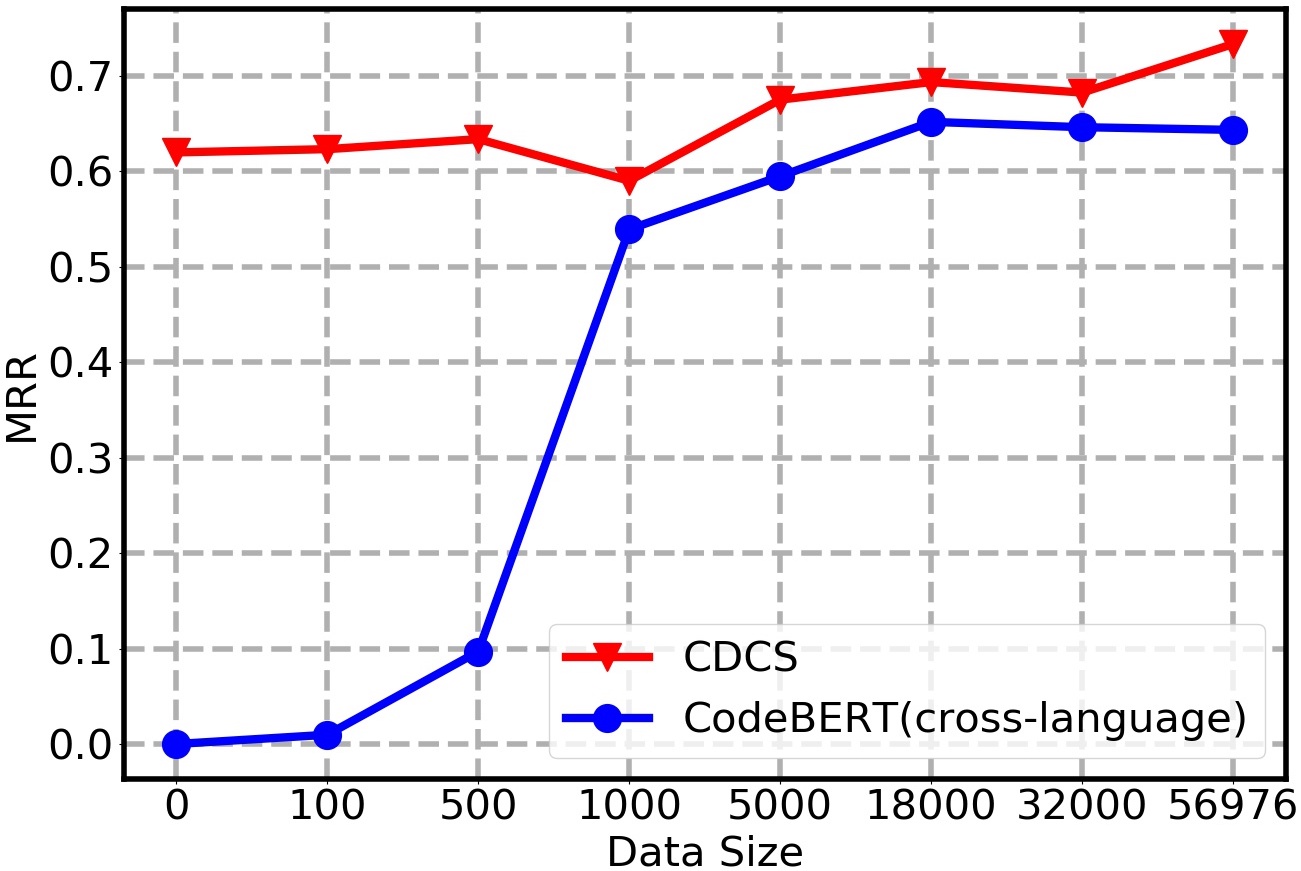}
        }
        \subfigure[Top-1 accuracy]{
            \includegraphics[scale = 0.092]{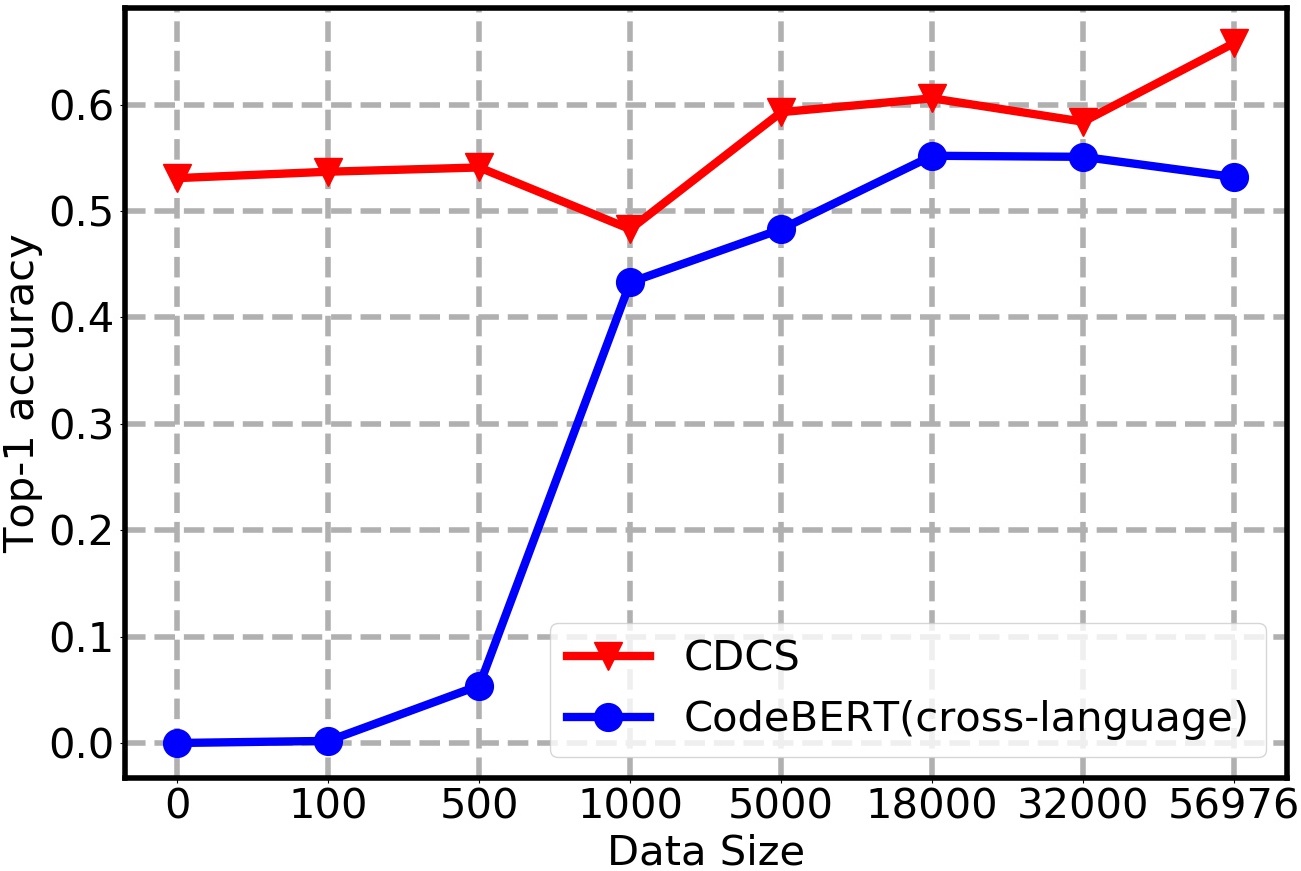}
        }
        \subfigure[Top-5 accuracy]{
            \includegraphics[scale = 0.092]{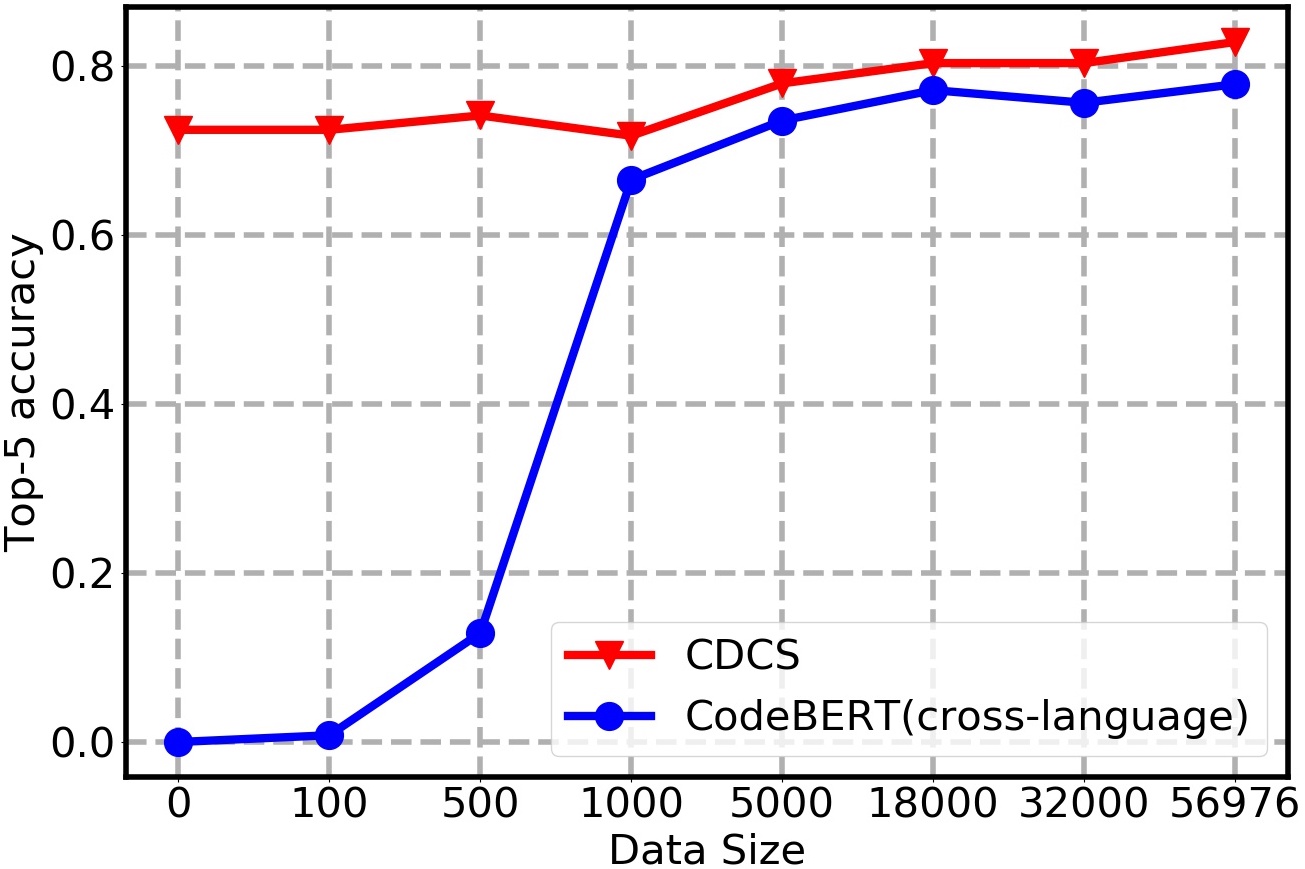}
        }
        \subfigure[Top-10 accuracy]{
            \includegraphics[scale = 0.092]{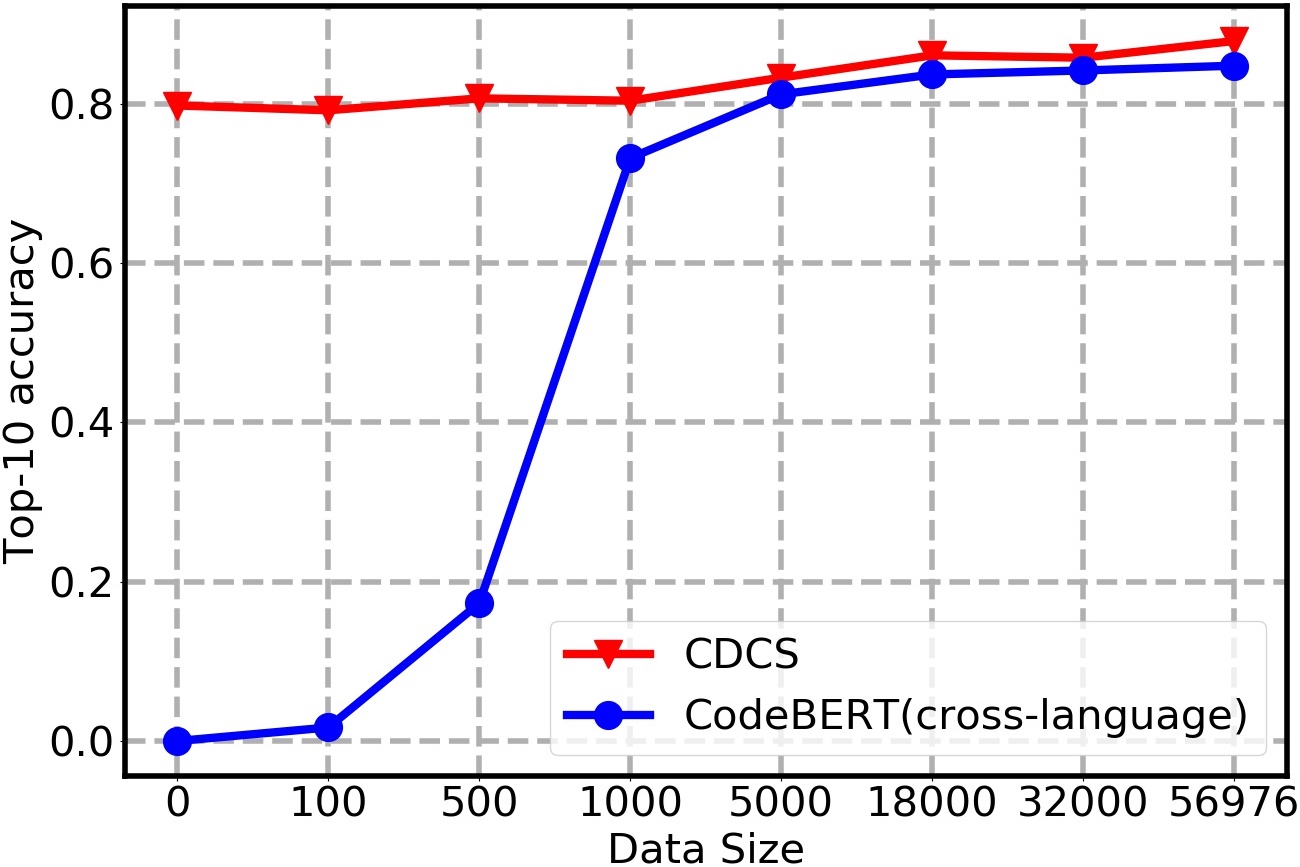}
        }
        \caption{Performance of \approach under different training data sizes on the Solidity dataset.}
        \label{RQ2_1_}
    \end{figure*}
	
	Overall, \approach achieves the best performance among all the methods.
	From the results on the SQL dataset, we can see that \approach outperforms the baseline models in terms of all metrics, especially the top-1 accuracy and MRR, which are about 11\% and 7\% greater than the strong baselines, respectively. 

	The improvement is more significant on the Solidity dataset (Table~\ref{table_solidity}). We can see that \approach substantially outperforms strong baselines especially in the top-1 accuracy and MRR, which are about 20\% and 18\% stronger, respectively.
	
	There is a large margin between CodeBERT (NL-based) and CodeBERT (within-domain). 
	We hypothesize that this is because the SQL corpus is too scarce, so that the pre-training may 
	not provide sufficient prior knowledge to the code-search model.
	\approach obtains more significant improvement against CodeBERT (NL-based) in SQL than that in the Solidity dataset, probably because SQL is much closer to natural language than Solidity.
	
	The results demonstrate that \approach is remarkably effective in domain-specific code search tasks.

    

\subsection{Effect of Data Size (RQ2)}
	
	Figure~\ref{RQ2_1} and \ref{RQ2_1_} show the performance of \approach under different data sizes compared with the cross-language CodeBERT~\cite{salza2021effectiveness}. 
 	We vary the size of training data from 0 to full data. 

	As the result shows, \approach outperforms the baseline model under all data sizes, which supports the significance of the improvement achieved by \approach. In particular, we note that when the data size gets smaller (e.g., $<$500), the improvement of \approach against the baseline model becomes more significant. That means that \approach is particularly effective in scarce data, indicating the outstanding ability of \approach on domain specific languages. By contrast, the baseline model without meta learning can not adapt to the task well due to the insufficiency of data.

	\subsection{Performance on other Pre-trained Models (RQ3)}
	\begin{table}[htbp]  
		\centering
		\small
		\caption{Performance of each method based on GPT-2.}
		\label{table_gpt2}
		\begin{tabular}{l@{}ll@{}l@{}l@{}l}  
			\toprule   
			Language\; & \;\;Model &  Acc@1 \; & Acc@5 \; & Acc@10\; &  MRR   \\  
			\midrule   
			\multirow{4}*{SQL} & No-Pretraining & 0.002 & 0.010 & 0.022 & 0.0124 \\
			& GPT2 (NL-based) & 0.481  & 0.808  & 0.889  & 0.6204 \\
			& GPT2 (within-domain) & 0.470 & 0.785 & 0.877 & 0.6088 \\
			& GPT2 (cross-language) & 0.447 & 0.767 & 0.875 & 0.5899 \\
			& $\mathrm{\approach}_\mathrm{GPT-2}$ & \bf 0.511 & \bf 0.823 &  \bf 0.905 & \bf 0.6464 \\ 
			\midrule   
		    \multirow{4}*{Solidity} & No-Pretraining & 0.002 & 0.008 & 0.014 & 0.0101 \\
		    & GPT2 (NL-based) & 0.484 & 0.751 & 0.830 & 0.6079 \\
			& GPT2 (within-domain) & 0.487 & 0.772 & 0.848 & 0.6073 \\
		    & GPT2 (cross-language) & 0.481 & 0.760 & 0.827 & 0.6057 \\
			& $\mathrm{\approach}_\mathrm{GPT-2}$ & \bf 0.561 & \bf 0.781 &  \bf 0.846 & \bf 0.6607 \\ 
			\bottomrule  
		\end{tabular}
	\end{table}	
	
	We evaluate the performance of $\mathrm{\approach}_\mathrm{GPT-2}$ and compare it with baseline models that are also based on GPT-2. 
	We experiment with $(Python, Java)$ as the source languages and test the performance in Solidity and SQL.
	The training differs a little bit in the meta learning phase: we formulate the input for code search as: $$[BOS], w_1,\dots, w_N, c_1,\ldots, c_m, [EOS]$$ where $[BOS]$ and $[EOS]$ represent the ``beginning'' and ``ending'' of the sequence, respectively. The representation of the $[EOS]$ token stands for the aggregated sequence representation and is used for classification. 
    We implement $\mathrm{\approach}_\mathrm{GPT}$ based on the Huggineface repository$^1$. The hyperparameters are set as follows: we set the batch size to 44, learning rate to 2.5$e$-4~\cite{radford2019language} which warms up in the first 1,000 steps and decays according to a cosine curve.
    
	Table~\ref{table_gpt2} shows the performance of $\mathrm{\approach}_\mathrm{GPT-2}$ compared against baseline models. Clearly, $\mathrm{\approach}_\mathrm{GPT-2}$ works better than all the baseline models. The MRR scores of $\mathrm{\approach}_\mathrm{GPT-2}$ are about 5\% and 10\% greater than those of the baseline model in the SQL and Solidity languages, respectively. This affirms the effectiveness of $\mathrm{\approach}_\mathrm{GPT-2}$ when equipped with meta learning.
	
	We notice that the GPT-2 pre-trained in natural language corpus shows a comparable performance to ours in the SQL language. We conjecture that SQL is simple and similar to natural languages, hence pre-training on massive text corpus is effective for the target task without heavy adaptation.
	Another notable point we observe is that the results of $\mathrm{\approach}_\mathrm{GPT-2}$ are lower than those of $\mathrm{\approach}_\mathrm{BERT}$, presumably because GPT-2 is a unidirectional language model, which dynamically estimates the probability of text sequences and can be more suitable for generation than search tasks. GPT-2 processes each input text from left to right sequentially, thus can be limited in representing context-sensitive features. By contrast, BERT-style models are trained with de-noising strategies (e.g., the MLM task) which enable them to obtain bidirectional, context-sensitive features.
	

	\begin{figure}[!tbp]
        \centering
        \subfigure[Batch sizes (SQL)]{
            \label{RQ4:a}
            \includegraphics[scale = 0.085]{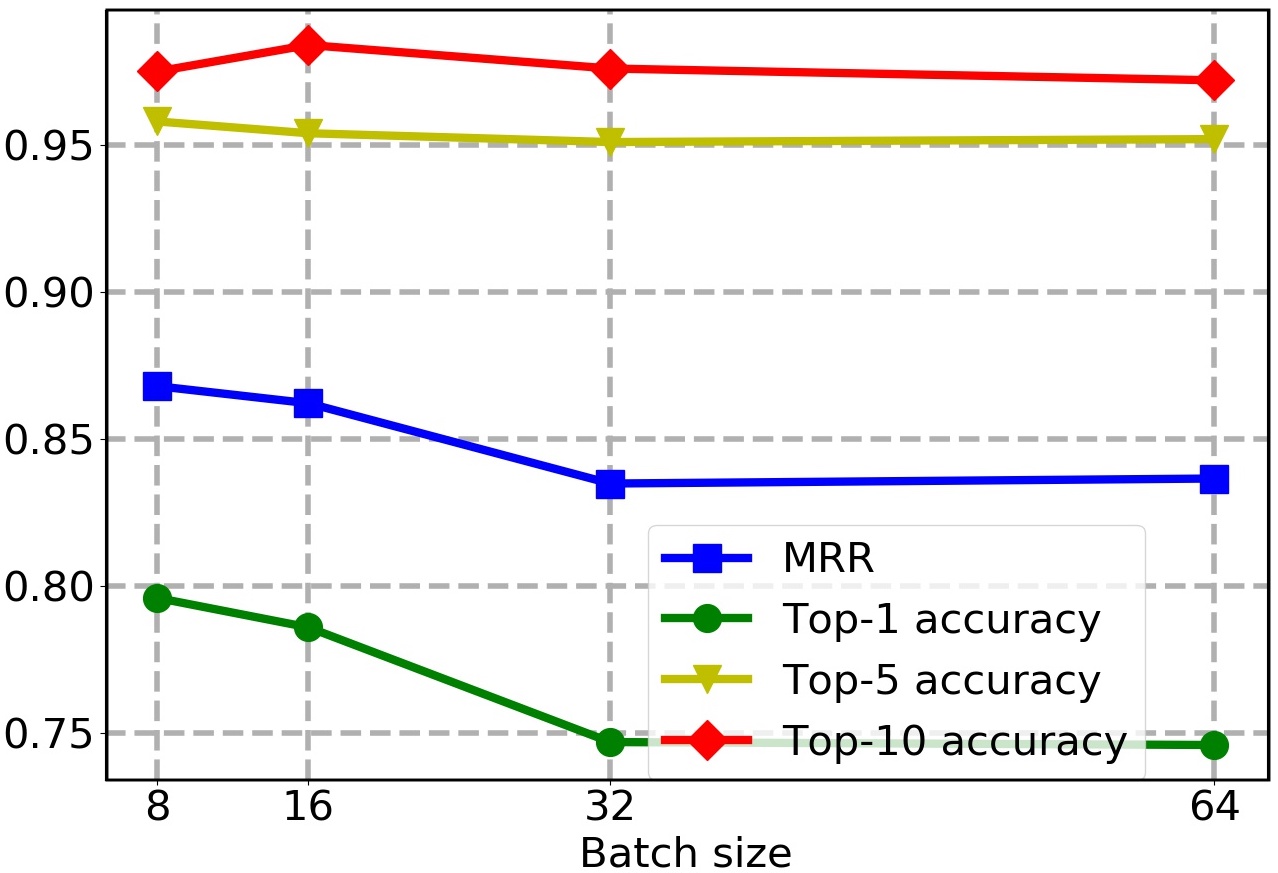}
        }
        \subfigure[Batch sizes (Solidity)]{
            \label{RQ4:b}
            \includegraphics[scale = 0.085]{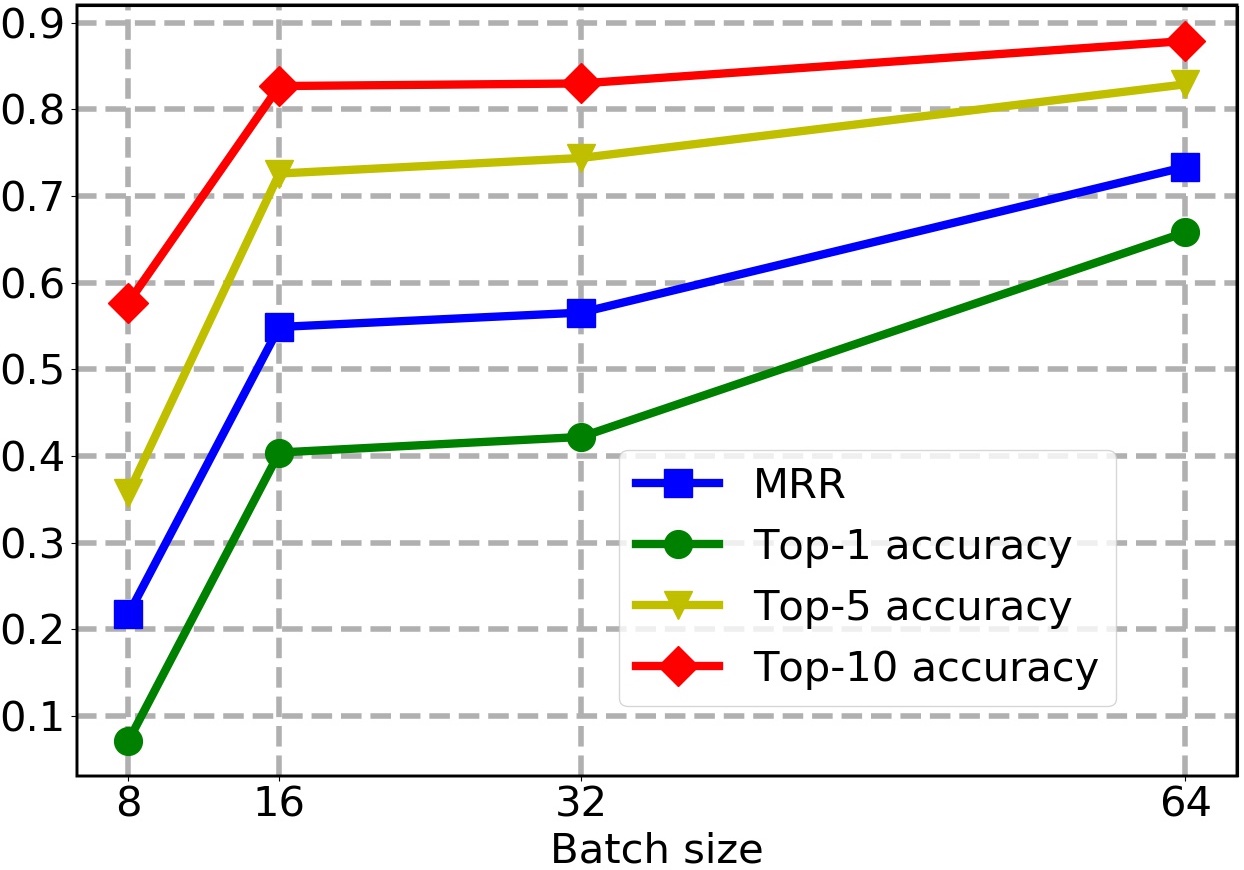}
        }

        \subfigure[Learning rates (SQL)]{
            \label{RQ4:c}
            \includegraphics[scale = 0.085]{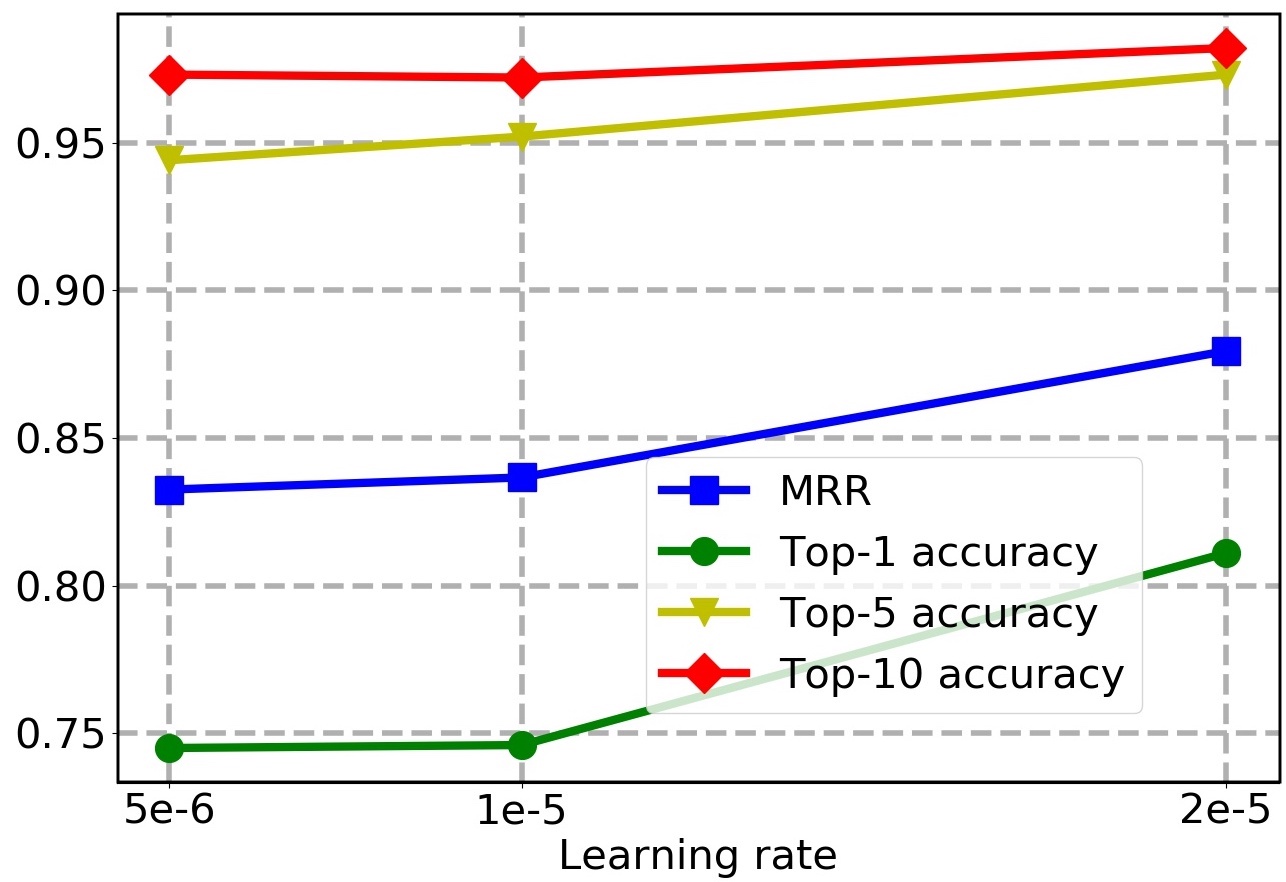}
        }
        \subfigure[Learning rates (Solidity)]{
            \label{RQ4:d}
            \includegraphics[scale = 0.085]{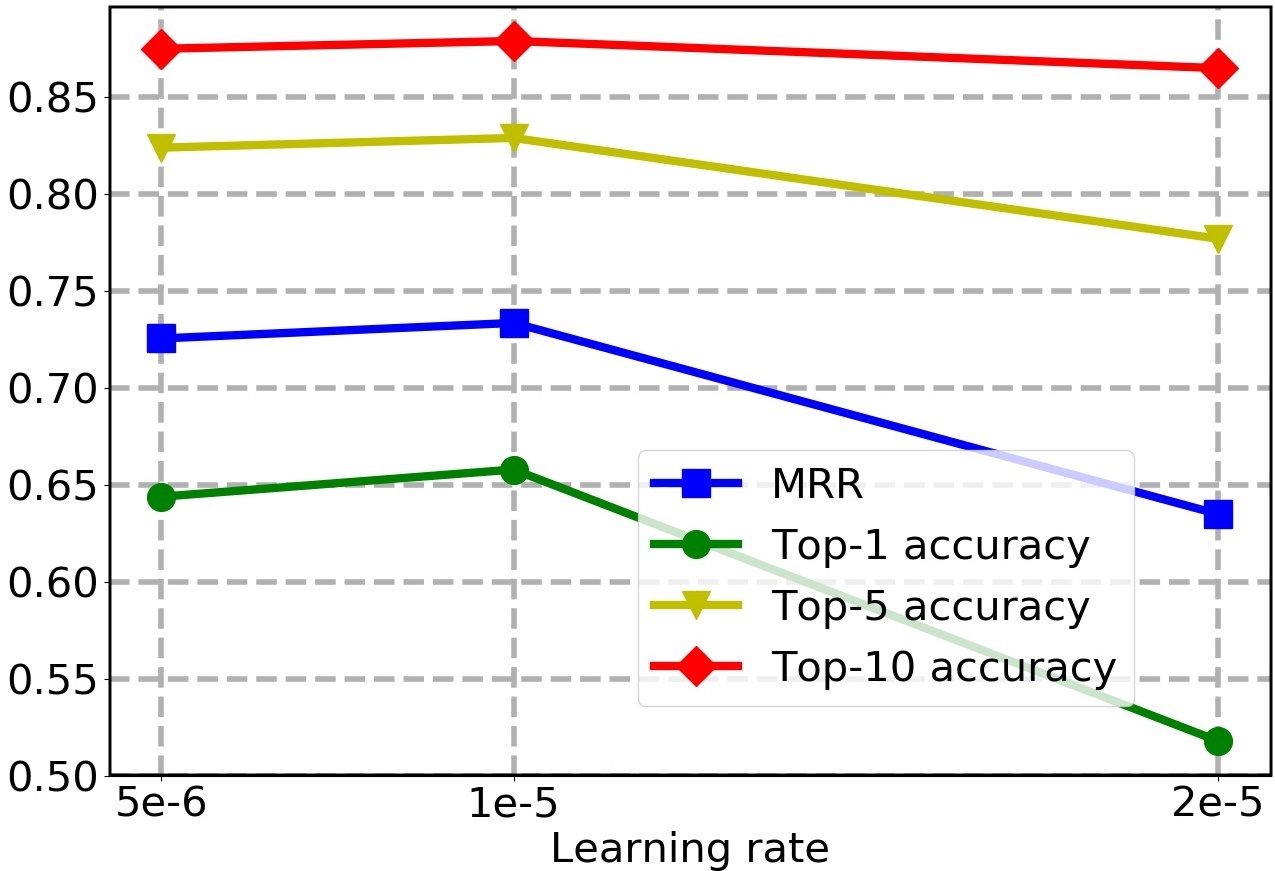}
        }
        \caption{Performance of \approach under different batch sizes (a-b) and learning rates (c-d).}
        \label{RQ4}
    \end{figure}
    
		\subsection{Impact of Different Hyperparameters (RQ4)}	
		
		Figure~\ref{RQ4:a} and \ref{RQ4:b} show the performance of \approach under different batch sizes on the SQL and Solidity datasets. We vary batch sizes to 64, 32, 16 and 8, respectively. The results show that larger batch sizes have slight impact on the performance, while smaller batch sizes have evident effect on the performance. 
		
		Figure~\ref{RQ4:c} and \ref{RQ4:d} show the performance of \approach under different learning rates on the SQL and Solidity datasets. We vary the learning rate to 2$e$-5, 1$e$-5, and 5$e$-6, respectively. As we can see, the performance is insensitive to learning rates lower than 1$e$-5. However, learning rates that are larger than 1$e$-5 have significant impacts on performance.
		
		To sum up, the impact of hyperparameters on \approach is limited to a certain range. The performance is sensitive to the hyperparameters when the batch size is less than 32 or the learning rate is greater than 1$e$-5. In addition, our model is more sensitive to both batch size and learning rate on the Solidity dataset than SQL. 

	\subsection{Case Study}
	We now provide specific search examples to demonstrate the effectiveness of \approach in domain specific code search.


	Listing 1 and 2 compare the top-1 results for the query ``what is the smallest city in the USA'' returned by \approach and the cross-language CodeBERT, respectively. The query involves complex semantics such as the word \texttt{smallest}. A code search system is expected to associate ``small'' with the corresponding SQL keyword \texttt{MIN}. They are different but are semantically relevant. Listing 1 shows that \approach can successfully understand the semantics of \texttt{smallest}, while the cross-language CodeBERT cannot. The example suggests that \approach is better than the cross-language CodeBERT~\cite{salza2021effectiveness} in terms of semantic understanding.
	
	Listing 3 and 4 show the results returned by \approach and the cross-language CodeBERT for the query ``Reset all the balances to 0 and the state to false'' in the Solidity language. The keywords in the query are \texttt{balances}, \texttt{state}, and \texttt{false}. It can be seen that both approaches return code snippets that hit some of the keywords. However, the snippet returned by \approach is clearly more relevant than that returned by the cross-language CodeBERT. For example, it explicitly states \texttt{benificiary.balance=0} and \texttt{filled = false} in the source code. On the other hand, the snippet provided by the cross-language CodeBERT is vague in semantics. Cross-language CodeBERT may pay more attention to similar words and is limited in understanding semantics. 

    These examples demonstrate the superiority of \approach in cross-domain code search, affirming the strong ability of learning representations at both token and semantic levels.




        \label{listing 1}
    	\begin{lstlisting}[language=SQL, caption={The first result of query "what is the smallest city in the USA" returned by \approach.}]
    SELECT  city_name
    FROM    city
    WHERE   population = ( 
                SELECT  MIN( population )
                FROM    city
            );
        \end{lstlisting} 
        
        \label{listing 2}
    	\begin{lstlisting}[language=SQL, caption={The first result of query "what is the smallest city in the USA" returned by the cross-language CodeBERT.}]
    SELECT  population
    FROM    city
    WHERE   population = ( 
                SELECT  MAX( population )
                FROM    city
            );
        \end{lstlisting} 
        
        \begin{lstlisting}[language=C++, caption={The first result of query "Reset all the balances to 0 and the state to false." returned by \approach.}]
    contract c8239{
        function clean() public
        onlyOwner {
            for (uint256 i = 0; i < addresses.length; i++)
            {
                Beneficiary storage beneficiary = beneficiaries[addresses[i]];
                beneficiary.balance = 0;
                beneficiary.airdrop = 0;
            }
            filled = false;
            airdropped = false;
            toVault = 0;
            emit Cleaned(addresses.length);
        }
    }
        \end{lstlisting} 
        
         \begin{lstlisting}[language=C++, caption={The first result of query "Reset all the balances to 0 and the state to false." returned by the cross-language CodeBERT.}]
   contract c281{
        function setTransferAgent(address addr, bool state) external onlyOwner inReleaseState(false) {
            transferAgents[addr] = state;
        }
    }
        \end{lstlisting} 
        
	    

	\subsection{Summary}
	Across all the experiments, the performance of the experimental group using pre-training is better than those without pre-training, and the evaluation results of the \approach experimental group combined with meta learning are better than those only trained with pre-training and fine-tuning. These results suggest that both transfer learning (pre-training \& fine-tuning) and meta learning have significant efficacy in deep code search. 

	The advantages of meta learning can be particularly seen from the experimental results of RQ2. The accuracy gap between \approach and the baseline models is becoming more significant as the data size decreases, which means that the size of training data has little effect on \approach. 
	Furthermore, the results of RQ3 suggest that our approach can be generalized to other pre-trained models such as GPT-2.
	
	Overall, the experimental results suggest that \approach has remarkable effectiveness in cross-domain code search especially when the training data is scarce.

    

 	

	\section{Discussion}

	\subsection{Why does \approach work better than the cross-language CodeBERT?}
	\label{ss:why}
\begin{figure*}[htbp]
        \includegraphics[scale = 0.4]{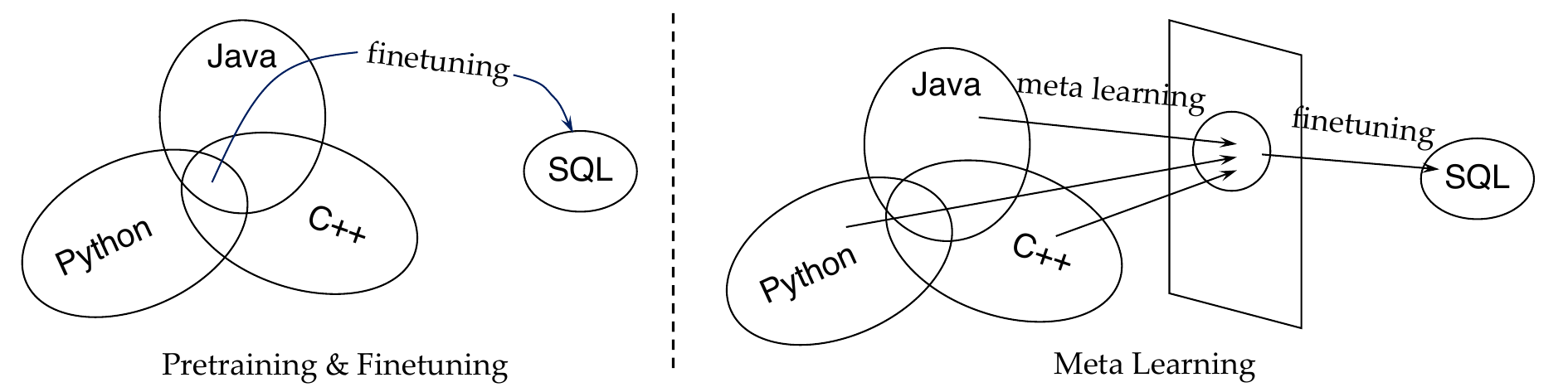}
        \caption{An illustration of the difference between meta learning and simply pre-training \& fine-tuning.}
        \label{fig5}
    \end{figure*}

	We believe that the advantage of \approach mainly comes from the difference between meta learning and simply pre-training \& fine tuning. As Figure~\ref{fig5} illustrates, the traditional \emph{pre-training \& fine-tuning} paradigm tries to learn the common features of multiple source languages in the pre-training phase, and directly reuses the pre-trained parameters to specific tasks through fine-tuning. The features of different source languages distract each other, leading to an ill-posed representation to be reused by the target language. By contrast, \emph{meta learning} employed by \approach tries to adapt the pre-trained parameters to new tasks during the learning process, resulting in representations that take into account all source languages.
    
	In a view of machine learning, both the \emph{pre-training \& fine-tuning} paradigm and \emph{meta learning} aim to enhance the generalization ability of deep neural networks in multiple tasks. However, in the \emph{pre-training \& fine-tuning} paradigm, the model will not obtain task information before fine-tuning on specific downstream tasks, while \emph{meta learning} focuses on learning information in specific tasks and can enhance the generalization ability of the model. \approach successfully combines the two methods.


	\subsection{Limitations}
    Although effective, we recognize that the adaptation of meta-learning to code search might not be a perfect fit. 
    Meta-learning is usually used for classification tasks on scarce data~\cite{yin2020metalearning,finn2017maml}, whereas we adapt it to the context of code search. These two concepts (i.e., classification vs. ranking) are not a natural fit. Hence, meta-learning might not perfectly solve the root problem of cross-domain code search. More adaptations are demanded to fit the two concepts. 
    
    In order to efficiently adapt code search tasks to scarce data scenarios, we follow the MAML paper~\cite{finn2017maml} and divide the data into machine learning ``tasks'', with each task aiming at training a code search model with small sized data. Such an approach has a few benefits. For example, it is easy for task adaptations since it does not introduce any learned parameters. Furthermore, adaptation can be performed with any amount of data since it aims at producing an optimal weight initialization~\cite{finn2017maml}.
    The limitation is that, the division of the data into ``tasks'' is random and there needs a concrete explanation on how split tasks are related to cross-language code search. It remains to investigate how such divisions turn out to be effective in scarce data.
    
    
	Another downside of \approach is that the MAML algorithm it employs can bring more time and computational cost in the large-scale data set. Different from the conventional gradient descent methods, MAML needs to compute a meta gradient based on multiple losses computed from sub-tasks. This costs extra time for saving model parameters and gathering meta gradients. 
	For example, in our experiments, it requires around 50\% extra hours for meta-learning compared to the baseline models.
	We leave more efficient transfer learning techniques for future directions.

	\subsection{Threats to Validity}
	
	We have identified the following threats to our approach:
     
    \textit{The number of source languages.}
    Due to the restriction of computational resources, we only selected two source languages and two domain-specific target languages. Meta learning with more source languages could have different results. In our future work, we will evaluate the effectiveness of our approach with more source and target languages.
    
	
	\textit{The selection of pre-training tasks.}
	The original CodeBERT uses two pre-training tasks, namely, masked language model (MLM) and replaced token detection (RTD)~\cite{feng2020codebert}. However, in our experiments, we only use the MLM as the pre-training task. Combining MLM with RTD may have effects on the results.
	 However, we believe that the results of the MLM task can stand for the performance of pre-training because the objective of RTD is similar to MLM in that both are based on the idea of de-noising. More importantly, the RTD task requires too much cost of time and computational resources, while the improvement it brings is marginal according to the ablation experiments in the CodeBERT paper~\cite{feng2020codebert}.
	 Moreover, compared with RTD, the MLM task is more widely used~\cite{wang2019bert} in domains other than programming languages.
	 
	 \textit{Generalization to other pre-trained models.}
	 We have built and evaluated our approach on top of two pre-trained models, namely, BERT and GPT-2. Thus, it remains to be verified whether or not the proposed approach is applicable to other pre-trained models such as BART~\cite{ahmad2021unified} and T5~\cite{mastropaolo2021t5code,wang2021codet5}.

	\section{Related Work}
	
	\subsection{Deep Learning Based Code Search}
	With the development of deep learning, there is a growing interest in adapting deep learning to code search~\cite{cambronero2019deep,gu2018deepcs,liu2020simplifying}. The main idea of deep learning based code search is to map natural and programming languages into high-dimensional vectors using bi-modal deep neural networks, and train the model to match code and natural language according to their vector similarities. 
	NCS (Neural Code Search)~\cite{sachdev2018retrieval} proposed by Facebook learns the embeddings of code using unsupervised neural networks.
	Gu et al.~\cite{gu2018deepcs} proposed CODEnn (Code-Description Embedding Neural Network), which learns the joint embedding of both code and natural language. CODEnn learns code representations by encoding three individual channels of source code, namely, method names, API sequences, and code tokens. UNIF~\cite{cambronero2019deep} developed by Facebook can be regarded as a supervised version of NCS. Similar to CODEnn, UNIF designs two embedding networks to encode natural and programming languages, respectively. Semantic Code Search (SCS)~\cite{husain2018create} first trains natural language embedding network and programming language embedding network respectively and then trains the code search task by integrating the two embedding network with similarity function. 
	CodeMatcher~\cite{liu2020simplifying}, which is inspired by DeepCS~\cite{gu2018deepcs}, combines query keywords with the original order and performs a fuzzy search on method names and bodies.
	Zhu et al.~\cite{zhu2020ocor} proposed OCoR, a code retriever that handles the overlaps between different names used by different developers (e.g., ``message'' and ``msg''). 
	Wang et al.~\cite{wang2022enriching} proposed to enrich query semantics for code search with reinforcement learning.
	
	While these methods are mainly designed for common languages, \approach focuses on domain-specific code search, where training data is often scarce and costly. \approach extends pre-trained models with meta learning to extract prior knowledge from popular common programming language for searching code written in domain-specific languages.
	
	\subsection{Pre-trained Language Models for Code}
	In recent years, pre-trained language models for source code have received much attention~\cite{feng2020codebert,phan2021cotext,ahmad2021unified,mastropaolo2021t5code}.
	CodeBERT~\cite{feng2020codebert}, built on top of the popular model of BERT~\cite{devlin2019bert}, is one of the earliest attempts that adapt pre-trained models for programming languages. CodeBERT is trained with six common programming languages (Python, Java, JavaScript, PHP, Ruby, and Go). Besides, they creatively proposed the replaced token detection (RTD) task for the pre-training of programming language.
	CoText~\cite{phan2021cotext} is a pre-trained Transformer model for both natural language and programming languages. It follows the encoder-decoder architecture proposed by~\cite{vaswani2017attention}. 
	PLBART~\cite{ahmad2021unified} learns multilingual representations of programming and natural language jointly. It extends the scope of pre-training to denoising pre-training, which involves token masking, deletion, and infilling. 
	Mastropaolo et al.~\cite{mastropaolo2021t5code} empirically investigated how T5 (Text-to-Text Transfer Transformer), one of the state-of-the-art PLMs in NLP, can be adapted to support code-related tasks. The authors pre-trained T5 using a dataset composed of English texts and source code, and then fine-tuned the model in four code-related tasks such as bug fix and code comment generation.

	Although these pre-trained models for source code can be used for cross-language code search~\cite{salza2021effectiveness} through pre-training in multiple languages and fine-tuning in the domain-specific language, they do not take into account the difference between source and target languages, and are limited in performing domain-specific code search. By contrast, \approach explicitly transfers representations of multiple source languages to the target language through meta learning.

	\subsection{Transfer Learning for Code Search}
	To our knowledge, there is only one previous work that is closely related to ours. Salza et al.~\cite{salza2021effectiveness} investigated the effectiveness of transfer learning for code search. They built a BERT-based model, which we refer to as \emph{cross-language CodeBERT}, to examine how BERT pre-trained on source code of multiple languages can be transferred to code search tasks of another language. Their results show that the pre-trained model performs better than those without pre-training, and transfer learning is particularly effective in cases where a large amount of data is available for pre-training while data for fine-tuning is insufficient~\cite{salza2021effectiveness}. 
	
	\approach differs significantly from theirs. We employ a meta learning algorithm to explicitly adapt the parameters from source languages to the target domain, while their work directly fine-tunes the pre-trained model in the target language. 
	
	
	


	\section{conclusion}
	
	In this paper, we present \approach, a cross-domain code search approach that reuses prior knowledge from large corpus of common languages to domain-specific languages such as SQL and Solidity. \approach extends pre-trained models such as CodeBERT with meta learning. 
	It employs a meta-learning algorithm named MAML which learns a good initialization of model parameters so that the model can quickly reach the optimal point in a new task with a few data samples.
    Experimental results show that \approach achieves significant improvement in domain-specific code search, compared to ``pre-training \& fine-tuning'' counterparts.
	In the future, we will investigate our method in more languages and other software engineering tasks.

    Source code and datasets to reproduce our work are available at:
    \url{https://github.com/fewshotcdcs/CDCS}
    .

	
	\section{Acknowledge}
     This work was sponsored by the National Natural Science Foundation of China under 62102244 and the CCF-Baidu Open Fund No.2021PP15002000. Xiaodong Gu is the corresponding author.
	
	\bibliographystyle{ACM-Reference-Format}
	\balance	
	\bibliography{references}
	\balance

\end{document}